\documentclass[lettersize,journal]{IEEEtran}
\usepackage{amsmath,amsfonts}
\usepackage{algorithmic}
\usepackage{algorithm}
\usepackage{array}

\usepackage[caption=false,font=normalsize,labelfont=rm,textfont=rm]{subfig}

\usepackage{textcomp}
\usepackage{stfloats}
\usepackage{url}
\usepackage[colorlinks,linkcolor=blue]{hyperref}
\usepackage{verbatim}
\usepackage{graphicx}
\usepackage{cite}
\usepackage{caption}
 \usepackage{booktabs}
\usepackage{url}
\usepackage{color}
\usepackage{mdframed}
\usepackage{amsmath}
\newtheorem{theorem}{Theorem}
\newtheorem{corollary}{Corollary}
\newtheorem{definition}{Definition}
\newtheorem{Lemma}{Lemma}

\newtheorem{assumption}{Assumption}

\newtheorem{remark}{Remark}

\usepackage{amssymb}
\usepackage{dsfont}

\def \ycrrr #1{\textcolor{black}{#1}} 
\def \yceee #1{\textcolor{black}{#1}} 
\def \ycbbb #1{\textcolor{black}{#1}} 
\def \ycaaa #1{\textcolor{black}{#1}} 
\def \ycccc #1{\textcolor{black}{#1}}

\begin{document}

\title{On  Random Sampling of Diffused Graph Signals with Sparse Inputs on Vertex  Domain}

\author{
  Yingcheng Lai,  \textit{Student Member,} Li Chai, \textit{Member, IEEE,} Jinming Xu, \textit{Member, IEEE}

\thanks{$^\dagger$ 
	Yingcheng Lai, Li Chai and Jinming  Xu are with the College of Control Science and Engineering, Zhejiang University, Hangzhou 310027, China. (Email: laiyingcheng@zju.edu.cn; chaili@zju.edu.cn; jimmyxu@zju.edu.cn). }%
}



\maketitle

\begin{abstract}
The sampling of graph signals has recently drawn much attention due to the wide applications of graph signal processing. While a lot of efficient methods and interesting results have been reported to the sampling of band-limited or smooth graph signals, few research has been devoted to non-smooth graph signals, especially to sparse graph signals, which are also of importance in many practical applications. This paper addresses the random sampling of non-smooth graph signals generated by diffusion of sparse inputs. We aim to present a solid theoretical analysis on the random sampling of diffused sparse graph signals, which can be parallel to that of band-limited graph signals, and thus present a sufficient condition to the number of samples ensuring the unique recovery for uniform  random sampling. Then, we focus on two classes of widely used binary graph models, and give explicit and tighter estimations on the sampling numbers ensuring unique recovery. We also propose an adaptive variable-density sampling strategy  to provide a better performance than uniform random sampling. Finally, simulation experiments are presented to validate the effectiveness of the theoretical results.
\end{abstract}

\begin{IEEEkeywords}
Graph signal processing, random sampling, interpolation, compressed sensing, sparse signal.
\end{IEEEkeywords}

\section{Introduction}
\IEEEPARstart{I}{n} recent years, graphs have been employed  as a robust framework for modelling the unstructured data and their intricate interactions over networks, and graph signal processing (GSP) has emerged as a powerful tool for analyzing  \yceee{data from different applications}\cite{2018Graph,2013The,2013Discrete}.
 The data on the graph might be personal hobbies, activity in brain regions, and traffic volume at stations.
 The extension of classical signal processing methods to graph signals is the goal of  GSP \cite{stankovic2019vertex,djuric2018cooperative}. Examples of relevant problems  include signal reconstruction\cite{2014Towards,2013The}, modeling and inference of diffusion processes\cite{difpro1,difpro2}, and topology recognition\cite{2013Discrete,toprec1}, etc.

The number of vertices  in real-world networks is typically enormous, such as urban temperature monitoring systems, social networks and neural brain networks.
Collecting
information from all vertices  in such large scale networks is  very difficult. 
 From this perspective, the sampling and recovery problem has received considerable attention as \yceee {one of the} cornerstone problems in the field of GSP  \cite{2013Signal, 2009Erratum,  2014Towards}. 
 Various results and methods have been reported for smooth graph signals, which have focused on the sampling and recovery problem  under the assumption that the graph signal is smooth, where graph signals have approximately band-limited characteristics on the graph Fourier basis\cite{2013Discrete,Tsitsvero2016Signals,2016Efficient,2015Random}. 
  In particular, ref. \cite{2009Erratum} makes  a pioneering contribution to the theory of GSP sampling and reconstruction, providing sufficient conditions to ensure unique recovery.
The necessary and sufficient conditions for unique recovery in undirected graphs were introduced in \cite{2014Towards}. Furthermore, the sampling results are extended to directed graphs \cite{2016Efficient,other2015Discrete}.
The above works discuss  the theoretical guarantee of unique recovery under smooth signals, while some works  design specific sampling strategies for sampling problems.
The sampling strategies for graph signals generally include deterministic and random sampling strategies. Currently, deterministic sampling strategies are mainly studied using greedy optimization  methods\cite{other2015Discrete,2016Efficient}. For the random  sampling strategy, ref. \cite{2015Random} proposes a variable density probabilistic sampling strategy for smooth graph signals. 


In contrast to  smooth graph signals,
other works consider non-smooth graph signals generated by localized diffusion of sparse inputs  in graph diffusion models\cite{sparseInput1,sparseInput2,aggre2}.  
This model has practical importance in many applications, including source identification of rumors and opinions  in social networks, inverse problems related to biological signals on graphs, and  estimation  of diffusion processes in multi-agent networks. 
The work \cite{sparseInput1} investigates the design of a recovery algorithm for sparse inputs  using a known graph diffusion model when sampling partially diffused sparse signals. Subsequently, ref.  \cite{aggre2}  extends  it to a special type of locally  aggregated graph signal and discusses  recovery methods for identifying sparse inputs.
When the graph diffusion model is unknown,   the blind identification problem is investigated in \cite{sparseInput2,Blind1} under the assumption that the diffused graph signals can be collected from all vertices. These methods adopt the  joint estimation of the graph diffusion model and sparse inputs.  

In this paper, we employ the random sampling strategy to reconstruct the unknown sparse inputs from partially observed diffused graph signals. 
For uniform random sampling, we present an estimation on the number of samples and prove that it can guarantee the unique recovery with high probability. We show how this number of samples relates directly to the property of the diffused graph filter and other factors. Based on this result, we obtain explicit and tight bounds for two particular graph models: Erdős-Rényi (ER) random and small-world networks. We also propose an adaptive random sampling strategy to improve the performance.

Compressed sensing  states that 
one can recover sparse signals from far fewer samples or measurements than traditional methods by a sampling mechanism satisfying the incoherence condition\cite{ICS1}. The focus  on frames for dictionaries and  the design of  random measurement matrices with  Gaussian distribution  has gradually become two popular research directions in compressed sensing\cite{ICS2,ICS3}. It has been widely applied in channel coding\cite{ICS4}, data compression\cite{ICS5}, magnetic resonance imaging (MRI) reconstruction\cite{mri}, and so on. 
 For the diffused sparse signal, the graph diffusion model cannot be considered as a frame,
 and the design of the sampling matrix cannot follow a Gaussian distribution either. 
Based on the above discussion, we have found that there are still some fundamental issues to be addressed in random sampling of non-smooth diffused graph signals with sparse inputs:
\begin{itemize}
    \item The theoretical guarantee for recovering uniqueness is not explicit. Specifically, how many vertices  need to be sampled for a given graph diffusion model to ensure accurate reconstruction of sparse inputs?
    
 \item Experiments show that compared to the ER random graph, the star-shaped graph and the regular graph exhibit poorer recovery performance for the same sample size. For ER random graphs with different connection probabilities, edges that are too  sparse or too  dense can lead to poor recovery. This naturally raises the question:  What is the relationship between the  samples  number for recovery  and the network  structure?  

   \item 
   Is there an alternative sampling strategy that can be applied to diffused sparse signals with higher recovery accuracy than  uniform  random sampling? 
\end{itemize}

 To address the above problems in this paper,  we derive the following new results:
\begin{itemize}
    \item \textbf{Sufficient conditions for ensuring the uniqueness of recovery}.  We present a sufficient condition for  the number of samples 
    ensuring the signal reconstruction under uniform  random sampling.  The result reveals that the number of samples is  related to  the incoherence parameter $\mu$, the sparse condition number ${\kappa (\Gamma)}$ induced by the diffusion matrix  and the sparsity $k$ of the sparse inputs.   Moreover, we show that the calculation  of ${\kappa (\Gamma)}$ can be simplified  when the elements in  the diffusion matrix and the sparse inputs are non-negative. 
    
        \item \textbf{Sampling performance analysis of typical networks}.   For binary graph  diffusion model, we  present the direct   relationship between the 
        number of samples ensuring recovery uniqueness and the network structure. 
          For ER random networks, approximately  $\sim \log n$ samples are sufficient for recovery uniqueness. Additionally, the relationship between the number of samples required and connection probability was revealed, and it was found that the required sample number reaches its minimum when the connection probability is 0.5.   For a small-world network of degree $d$,  we characterize  the relationship between the number of samples required and the adjacency matrix of the $d$-regular graph.  We show that the number of samples needed decreases as the probability of rewiring increases. 

    \item \textbf{New  sampling strategy}.   An  alternative
     sampling strategy is proposed to  exploit  variable density sampling techniques. 
     We  prove that the variable  density sampling strategy  offers a performance guarantee  with fewer  samples  than  uniform random sampling. 
\end{itemize}
\textbf{Notations}.   $X^*$ represents the conjugate  transpose of $X$. $\left\| x \right\|_0$ indicates the $L_0$-norm, i.e., the number of non-zero elements in the $x$.  ${\left\| x \right\|_1} = \sum {\left| {{x_i}} \right|} $ and  ${\left\| x \right\|_1} = \sum {\left| {{x_i}} \right|} $ 
is the $L_1$-norm, and $\left\| X \right\|_2$
denotes  the standard Euclidean norm. $\left\| x \right\|_\infty$ represents the maximum absolute value of the vector $x$, and $\left\| X \right\|_\infty$ represents the maximum value in the sum of absolute values for each row of the matrix $X$.
 We use $\left\| X \right\|_{1 \to 2}$ to represent the largest $L_2$-norm among all columns of $X$. 
   $ {\left\| X \right\|_{1,1}} = \sum\limits_{i,j} {\left| {{x_{ij}}} \right|} $   denotes  the sum of absolute values of matrix elements of $X$.
  $I$ and $\mathbf{1}$ denote the identity matrix and the all-one  vector of appropriate size, respectively.
\section{problem Formulation}

\subsection{Network  and  Diffusion Model }

\yceee{An undirected graph with $n$ vertices is denoted as}   $\mathcal{G} = (\mathcal{N}, \mathcal{E}, A)$, where $\mathcal{N}$ is the set of vertices, $\mathcal{E}$ is the set of edges, and a pair $(i, j) \in \mathcal{E}$ indicates that there is a link between vertex $i$ and vertex $j$. 
The matrix $A \in \mathbb{R}^{n \times n}$ denotes the adjacency matrix, where the  $a_{i,j}$ represents the weight of the relationship between vertices  $i$ and $j$.  The degree of vertex $i$ \yceee{is defined as}  ${d_i} = \sum\nolimits_{j = 1}^n {{a_{ij}}}$. 
Associated with the given graph $\mathcal{G}$, a graph signal is represented as \yceee{$x = [x_1, x_2, \ldots, x_n]^T \in \mathbb{R}^n$,} where the $i$-th element $x_i$ represents the signal value at vertex $i$. 

In this paper, we consider the following diffused graph signal model
\begin{equation} \label{1.2}
 x = H\alpha ,
 \end{equation}
 where $H$ is the graph diffusion matrix, and $ \alpha  \in {\mathbb{R}^n} $ is the sparse inputs with sparsity $k$, i.e. ${\left\| \alpha  \right\|_0}  \le  k$.
Our goal is to  recover sparse inputs  $\alpha $ by randomly sampling certain number  of vertices,  and to explore the relationship between the number of samples ensuring unique recovery and the  network structure. 

\subsection{ Problem Formulation}

\ycrrr{Let $\mathcal{M} = \{\omega_1, \ldots, \omega_m\}$  be the sampling set, where $m= \left| \mathcal{M} \right| $ is the number of samples and ${\omega _i \in {1,\ldots, n}}$. 
Then, the sampling matrix ${C_\mathcal{M}} \in {\left\{ {0,1} \right\}^{m \times n}}$ is defined as}
\begin{equation}\label{1.7}
 \left[ {{C_\mathcal{M}}} \right] _{i,j}
  \buildrel \Delta \over =  
\left\{ {\begin{array}{*{20}{c}}
  1,&{{\text{if }}j = {\omega _i}}, \\ 
  0,&{{\text{otherwise}}} .
\end{array}} \right.\end{equation}
To represent the  sampling process, we denote the probability distribution over $\{   1, \ldots, n\}$ by $\mathcal{P}$, and the sampling probabilities of all vertices  by \yceee{$p = \left[ p_1, \ldots, p_n \right] \in \mathbb{R}^n$} which  satisfies the constraint $\sum_{i=1}^n p_i = 1$. 
 The sampling set $\mathcal{M} $ is formed by independently selecting $m$ indices from the vertex set according to the probability distribution $p$, i.e.,
\begin{equation}\label{1.5}
    \mathbb{P} \left( {{\omega _j} = i} \right) = {p_i}, \quad{\rm{  }}\forall j \in \left\{ {1,...,m} \right\},  \forall i \in \left\{ {1,...,n} \right\}.
\end{equation}

For the uniform  random sampling processes, the probability of sampling each vertex is identical, i.e., $p = [1/n,...,1/n]$. Note that the sampling procedure permits duplicate vertices  in the sampling set $\mathcal{M}$. In practical engineering, each selected vertex can only be sampled once, with duplicates added artificially afterwards.

\begin{figure}[H]
\centering
	\includegraphics[width=0.8\linewidth]{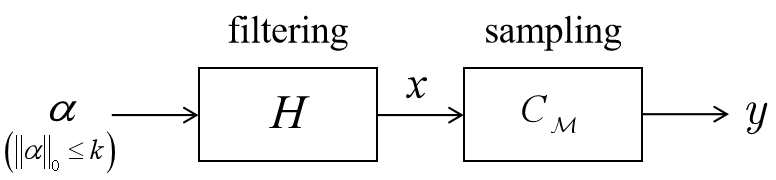}
	\caption{The generation and sampling process of diffused sparse signals.}
 \label{liucheng}
\end{figure}

With the  sampling matrix ${C_\mathcal{M}}$, the observed signal $ y \in {\mathbb{R}^m} $ can be formulated  as: 
\begin{equation} \label{the3.1}
\begin{aligned}
 y = {C_\mathcal{M}}x = {C_\mathcal{M}}H\alpha =  H_\mathcal{M} \alpha,  
\end{aligned}
\end{equation}
 where  ${H_\mathcal{M}} \buildrel \Delta \over =  {C_\mathcal{M}}H$. The  system  \eqref{the3.1} is shown by Fig. \ref{liucheng}.
 
It is known from linear algebra that the reconstructed signal $\alpha$ has an infinite number of  solutions when $m \ll n$. However, compressed sensing theory  states that the uniqueness of the solution can be guaranteed  when  $\alpha$ is a sparse vector. It is typically relaxed to a convex optimization problem as follows \cite{orth} 
\begin{equation} \label{the3.3}
    \begin{array}{*{20}{c}}
  {({P_1})\;}&{\begin{array}{*{20}{c}}
  {{\text{    }}\mathop {\min }\limits_\alpha  {\text{ }}{{\left\| \alpha  \right\|}_1}} \\ 
  {{\text{s}}.{\text{t}}.\;{\text{   }}y = {H_\mathcal{M}}\alpha ,} 
\end{array}} 
\end{array}{\kern 1pt} \;
\end{equation}
  Various sufficient conditions have been proposed in the literature to  accurately recover $\alpha$ in
 compressed sensing\cite{ICS4,ICS3}. A metric condition called Restricted Isometric Property (RIP) is introduced in  \cite{ICS4}. Specifically, it stated that the matrix $H_\mathcal{M}$ satisfies the RIP of order $k$ if 
 \begin{equation}
    \left( {1 - \delta } \right){\left\| \alpha  \right\|_2} \le {\left\| {{H_\mathcal{M}}\alpha } \right\|_2} \le \left( {1 + \delta } \right){\left\| \alpha  \right\|_2}  \nonumber
\end{equation}
 holds  for all $k$-sparse vector $\alpha$  with a small positive  constant $\delta$.  Some random matrices  satisfy  this property, such as matrix's elements obeying Gaussian or Bernoulli random distributions\cite{ICS3}. In addition, the mutual incoherence  can also guarantee the recovery of 
 $\alpha$ \cite{mutualincoherence1,mutualincoherence2}, which requires the matrix ${{H_\mathcal{M}}}$ to be approximately orthogonal, i.e.,  
  \begin{equation}
  \begin{aligned}
       \mu \left( {{H_\mathcal{M}}} \right) < {1 \mathord{\left/
 {\vphantom {1 {\left( {2k - 1} \right)}}} \right.
 \kern-\nulldelimiterspace} {\left( {2k - 1} \right)}}, \nonumber
  \end{aligned}
 \end{equation}
 where $\mu \left( {{H_\mathcal{M}}} \right) \triangleq \mathop {\max }\limits_{i \ne j} \frac{{\left| {h_i^T{h_j}} \right|}}{{{{\left\| {{h_i}} \right\|}_2}{{\left\| {{h_j}} \right\|}_2}}}$ .
 

\section{Main Result}
In this section,  we will establish a sufficient condition for the number of samples to ensure unique recovery of $\alpha$ in  the problem $(P_{1})$ under uniform  random sampling.
First, we make the following assumption.

\begin{assumption} \label{ass1}
 Assume that the expectation of   matrix $ \mathbb{E} {{H_{\mathcal{M}}^*}}{H_{\mathcal{M}} } $  is invertible.
\end{assumption}

  With the Assumption \ref{ass1}, we define
\begin{equation} \label{Gamma}
    \Gamma  \triangleq m{\left[ {\mathbb{E}H_\mathcal{M}^*{H_\mathcal{M}}} \right]^{ - 1}}.
\end{equation}
 
 The number of measurements required to stably sample a diffused sparse signal will depend on a quantity called incoherence. We need the following definitions: 

\begin{definition}[Incoherence\cite{orth}]
The incoherence parameter is the smallest number ${\mu} $ such that
\begin{equation} \label{defi}
  \mathop {  \max }\limits_{1 \le  i,j \le  n} \left\{ {\left| {h_{i,j}^{}} \right|} \right\} \le  {\mu}  ,{\text{          }}  \mathop {\max }\limits_{1 \le  i,j \le n } \left\{ {\left| {\left[ {H\Gamma } \right]_{i,j}^{}} \right|} \right\} \le  {\mu},
\end{equation}
\end{definition}
where $h_{i,j}$ is the $(i,j)$-th element of the matrix $H$.

\begin{definition}[Sparse Condition Number\cite{kueng2014ripless}] \label{SparseNumber}
     The largest and smallest $k$-sparse eigenvalue of a matrix $X$ are given by
\begin{equation} \label{defcon.1}
\begin{aligned}
{\lambda _{\max }}\left( {k,X} \right)  \buildrel \Delta \over = & \mathop {\max }\limits_{ {{\left\| v \right\|}_0} \le  k} \frac{{{{\left\| {Xv} \right\|}_2}}}{{{{\left\| v \right\|}_2}}},\\
{\text{    }}{\lambda _{\min }}\left( {k,X} \right)  \buildrel \Delta \over = & \mathop {\min }\limits_{ {{\left\| v \right\|}_0} \le  k} \frac{{{{\left\| {Xv} \right\|}_2}}}{{{{\left\| v \right\|}_2}}}.
 \end{aligned}   \end{equation}
The $k$-sparse condition number  of $X$ is
\begin{equation} \label{defcon.2}
\begin{aligned}
\text{ cond}\left( {k,X} \right){\buildrel \Delta \over =}\frac{{{\lambda _{\max }}\left( {k,X} \right)}}{{{\text{ }}{\lambda _{\min }}\left( {k,X} \right)}}.
 \end{aligned}   \end{equation}
\end{definition}
Denote ${\kappa  }(X)$ as follows
   \begin{equation} \label{mR.2}
\begin{aligned}
  {\kappa  }(X) {\buildrel \Delta \over =} \max \left\{ {\text{cond}\left( {k,X } \right),\text{cond}\left( {k,{X ^{ - 1}}} \right)} \right\} .
   \end{aligned} 
  \end{equation}
  
 Now we can report the main result of this paper. In Theorem \ref{mainResult}, we provide a sufficient condition for the exact recovery of  $\alpha$  by the problem ($P_1$) when  uniform  random sampling. 

\begin{theorem} \label{mainResult}
Consider the system \eqref{the3.1}, where $\alpha$ is the $k$-sparse input, $H$ the graph diffusion matrix, and $ \mathcal{M} = \{\omega_1, \ldots, \omega_m\}$ the sampling set with $\omega_i$  being independent random variables with uniform distribution. With Assumption \ref{ass1} define the matrix $\Gamma$ as \eqref{Gamma}.
Then  with probability $ 1 - {e^{ - \varepsilon  }} - {3 \mathord{\left/
 {\vphantom {3 { n}}} \right.
 \kern-\nulldelimiterspace} { n}} $, the problem ($P_1$) has a unique minimizer  provided that
 \begin{equation} \label{mR.1}
\begin{aligned}
 m \ge  C\left( {1 + \varepsilon  } \right){\mu}^2 k  {\kappa (\Gamma)} \left( {\log n + \log {\mu} } \right) ,
 \end{aligned} 
  \end{equation}
  where  $C>0$ and $\varepsilon  >0$ are positive constants.
  \end{theorem}

The proof  of Theorem \ref{mainResult} is given in Section VI.

\begin{remark}
 From \eqref{mR.1}, it can be seen that the sampling boundary that guarantees unique recovery is related to the  incoherence parameter $\mu$,  the sparse condition number ${\kappa (\Gamma)}$, the cardinality of sparse inputs   $k$ and the 
 number of vertices $n$. Given a graph diffusion model $H$, we can calculate the required number of samples based on \eqref{mR.1}, which provides a theoretical guide for sampling. When $\mu$, ${\kappa (\Gamma)}$ and $k$ are small, it can be observed that we only need an order of magnitude proportional to $\log n$ to achieve unique recovery. 
\end{remark}

\begin{remark}
    The proof of Theorem \ref{mainResult} relies  on principles from compressed  sensing. 
     Similar work \cite{orth} considers the identical sampling model as ours, where $H$ consists  of the product of two orthogonal basis, and the result shows  that $m  \ge  C{{\mu}^2  } k\log (n/\delta)$, where  $\mu  = \mathop {\max }\limits_{1 < i,j < n} \left\{ {\left| {{h_{i,j}}} \right|} \right\} $.  
    From our result (Theorem \ref{mainResult}), it can be observed that when $H$ is a  matrix composed of the product of two orthogonal basis, the result is consistent with the work \cite{orth}.  
    
    Ref. \cite{unified} addresses the  compressed  sensing based on redundant dictionaries within an analysis sparsity model. 
    The analysis sparsity model originates from the sparse representation in harmonic analysis, assuming that the signal is sparse after transformation. 
    In contrast, we focus on the synthesis sparsity model, 
    where the  signal is a linear combination of a few atoms in the dictionary. In addition, according to the definition of our sparsity condition number, our result  reduces  the lower bound of the number of samples compared to \cite{unified}. 
    
 The work \cite{2015Random} can be regarded as a particular instance of the ref. \cite{unified} in smooth graph signals, where  signals exhibit  sparse features by graph Fourier transform. Specifically, according to the definition of smooth graph signals, the energy of signals in the graph Fourier domain is mainly concentrated in the first $k$ low frequencies\cite{2013Signal}. Therefore, the sparse support on graph Fourier domain is known.
In comparison to the work 
 \cite{2015Random}, our work considers  a more complex problem, in which the graph signal is characterised by sparsity on the vertex domain and the sparse support on the vertex domain  is unknown.

For the purpose of simplifying calculations,  we show that the $k$-sparse condition number of ${\left[ {\mathbb{E}H_\mathcal{M}^*{H_\mathcal{M}}} \right]^{ - 1}}$ will not be greater than the the $k$-sparse condition number of ${  {\mathbb{E}H_\mathcal{M}^*{H_\mathcal{M}}}   }$ if  $H$ is a non-negative matrix and $\alpha$ is a non-negative vector in the following corollary.

\begin{corollary} \label{lemm.sn}
With the same assumption of Theorem \ref{mainResult}, if all elements of $H$ are non-negative, the following inequality holds
	\begin{equation} \label{examsw.10}
		\begin{aligned}
			{\text{cond}}\left( {k,{\left[ {\mathbb{E}H_\mathcal{M}^*{H_\mathcal{M}}} \right]^{ - 1}} }\right) \le  {\text{cond}}\left( {k,{  {\mathbb{E}H_\mathcal{M}^*{H_\mathcal{M}}}  } }\right).
		\end{aligned} 
	\end{equation}
    Moreover, if all elements of $\alpha$ are non-negative, then with probability $ 1 - {e^{ - \varepsilon  }} - {3 \mathord{\left/
			{\vphantom {3 { n}}} \right.
			\kern-\nulldelimiterspace} { n}} $, the problem ($P_1$) has a unique minimizer provided that
	\begin{equation} \label{lemmsn.1}
		\begin{aligned}
		 m \ge  C\left( {1 + \varepsilon } \right){\mu ^2}k \cdot {\text{cond}}\left( {k, {\mathbb{E}H_\mathcal{M}^*{H_\mathcal{M}}} } \right) \cdot \left( {\log n + \log \mu } \right), 
		\end{aligned} 
	\end{equation}
	where  $C>0$ and $\varepsilon  >0$ are positive constants.
\end{corollary}

\emph{Proof:}
Let us assume that  $\beta =   \mathbb{E}H_\mathcal{M}^*{H_\mathcal{M}} \alpha $, then the maximum  $k$-sparse eigenvalue of ${{\left[ {\mathbb{E}H_\mathcal{M}^*{H_\mathcal{M}}} \right]}^{ - 1}}$ can be expressed as 
\begin{equation} \label{examsw.8}
	\begin{aligned}
		&{\lambda _{\max }}\left( {k,{{\left[ {\mathbb{E}H_\mathcal{M}^*{H_\mathcal{M}}} \right]}^{ - 1}}} \right) =  \mathop {\max }\limits_{{{\left\| \beta \right\|}_0} \le  k} \frac{{{{\left\| {{{\left[ {\mathbb{E}H_\mathcal{M}^*{H_\mathcal{M}}} \right]}^{ - 1}} \beta } \right\|}_2}}}{{{{\left\|  \beta  \right\|}_2}}} \\
		&=  \mathop {\max }\limits_{{{\left\|  \beta  \right\|}_0} \le  k} \frac{{{{\left\| \alpha \right\|}_2}}}{{{{\left\| {\mathbb{E}H_\mathcal{M}^*{H_\mathcal{M}} \alpha} \right\|}_2}}} 
		=    \frac{1}{{\mathop {\min }\limits_{{{\left\| \beta  \right\|}_0} \le  k} \frac{{{{\left\| {\mathbb{E}H_\mathcal{M}^*{H_\mathcal{M}}\alpha } \right\|}_2}}}{{{{\left\| \alpha  \right\|}_2}}}}}  .
	\end{aligned} 
\end{equation}
Similarly,  the minimum  $k$-sparse eigenvalue of ${{\left[ {\mathbb{E}H_\mathcal{M}^*{H_\mathcal{M}}} \right]}^{ - 1}}$ is
\begin{equation} \label{examsw.9}
	\begin{aligned}
		{\lambda _{\min }}\left( {k,{{\left[ {\mathbb{E}H_\mathcal{M}^*{H_\mathcal{M}}} \right]}^{ - 1}}} \right) =  \frac{1}{{\mathop {\max }\limits_{{{\left\| \beta  \right\|}_0} \le  k} \frac{{{{\left\| {\mathbb{E}H_\mathcal{M}^*{H_\mathcal{M}}\alpha } \right\|}_2}}}{{{{\left\| \alpha  \right\|}_2}}}}} . 
	\end{aligned} 
\end{equation}
Combining \eqref{examsw.8} and \eqref{examsw.9},  the $k$-sparse condition number of ${{\left[ {\mathbb{E}H_\mathcal{M}^*{H_\mathcal{M}}} \right]}^{ - 1}}$ can be expressed as
 \begin{equation}  
	\begin{aligned}
		& {\text{cond}}\left( {k,{{\left[ {\mathbb{E}H_\mathcal{M}^*{H_\mathcal{M}}} \right]}^{ - 1}}} \right)  
		=  \frac{{\mathop {\max }\limits_{{{\left\| \beta  \right\|}_0} \le  k} \frac{{{{\left\| {\mathbb{E}H_\mathcal{M}^*{H_\mathcal{M}}\alpha } \right\|}_2}}}{{{{\left\| \alpha  \right\|}_2}}}}}{{\mathop {\min }\limits_{{{\left\| \beta  \right\|}_0} \le  k} \frac{{{{\left\| {\mathbb{E}H_\mathcal{M}^*{H_\mathcal{M}}\alpha } \right\|}_2}}}{{{{\left\| \alpha  \right\|}_2}}}}}\\
        & \le   {\text{cond}}\left( {k,\mathbb{E}H_\mathcal{M}^*{H_\mathcal{M}}} \right)	, \nonumber
		\end{aligned} 
	\end{equation}
where the inequality  holds comes from that
each element of the matrix $\mathbb{E}H_\mathcal{M}^*{H_\mathcal{M}}$ and the vector $\alpha$ is non-negative, then  we have  ${\left\| \alpha  \right\|_0} \le  {\left\| \beta  \right\|_0} \le  k$. According to the definition of the $k$-sparse condition number in Definition \ref{SparseNumber}, we know that the sparse condition number increases monotonically as $k$ increases, so the inequality holds.
This completes the proof. 
$\hfill\blacksquare$


\end{remark}


\section{Binary Diffusion Model}
In this section, we analyse the sampling rate guaranteed to uniquely recover the problem ($P_1$) under a type of binary  diffusion model based on the results in Theorem \ref{mainResult}.
Specifically, we consider the case of  $\{0,1\}$-binary network diffusion, i.e., $H  =I+\delta A$, where $A$ is the unweighted adjacency matrix and $0<\delta \le 1$. This model is of practical interest in a number of applications, including  
the spread of epidemics and the spread of rumors in social networks and brain
networks, such as the Susceptible-Infected (SI) model\cite{Blind1}.

For example, the social interaction between people can be modeled as an undirected, symmetric sparse graph, and the spread of epidemics can  either be influenced or not. In this scenario, the impact of a vertex on its neighbors can be represented in a binary format with two possible values, namely 0 and 1. The initial number of epidemic sources initiated by a small group can be represented by $ \alpha $, where the support of non-zero elements represents the source of epidemics. 
Suppose that only part of the population of a given city is initially infected, and that the adjacency matrix is constructed on the basis of the daily social interactions between individuals.  It would be desirable to identify the source of the epidemic with minimal delay. However, it is impractical to test all individuals  for  epidemics  within a single day. Therefore, based on a  preliminary understanding of the social habits of the population in the city, we hope to estimate how many people need to be tested to find  the source of the epidemic.

In the following,  we will explore two particular types of graphs. 
First, considering the ER random graph, we demonstrate how connection probability $b$ affects the number of samples to reconstruct a sparse signal, i.e.,   $m \sim {{ - \log \left( {b - {b^2}} \right)} \mathord{\left/
 {\vphantom {{ - \log \left( {b - {b^2}} \right)} {\left( {b - {b^2}} \right)}}} \right.
 \kern-\nulldelimiterspace} {\left( {b - {b^2}} \right)}}$. For an appropriate connection probability $b$ (without an excessive bias towards $0$ or $1$), approximately  $ \sim \log n$ samples are sufficient for the problem ($P_1$).
 We then  investigate the number of samples for recovery of small-world networks and characterize  the relationship between the number of samples and the adjacency matrix of the $d$-regular graph. Furthermore, it is demonstrated  that  the number of samples required decreases as the rewiring probability $b$ increases. 

\subsection{Erdős-Rényi Random Network}
\begin{theorem}
\label{Lemma.er}
 For an ER random network with $n$ vertices  and edge connection probability $b$. Let $\alpha$ be a $k$-sparse vector diffused in the network by the binary  diffusion model $H = I+\delta A$, where $A$ is the  unweighted adjacency matrix and $ 0<\delta   \le  1 $.
Let $ \mathcal{M}$ be the sampling set as in Theorem \ref{mainResult}.   
Then with probability $ 1 - {e^{ - \varepsilon  }} - {3 \mathord{\left/
 {\vphantom {3 { n}}} \right.
 \kern-\nulldelimiterspace} { n}} $, the problem ($P_1$) has a unique minimizer  provided that
  \begin{equation} \label{erlem.1ll}
\begin{aligned}
  m \ge  \frac{{C\left( {1   + \varepsilon  } \right)k^{3/2} \left( {\log n - \log { {{\delta ^2}\left( {b - {b^2}} \right)} }  } \right)}}{{  {{\delta ^2}\left( {b - {b^2}} \right)}  }} ,
 \end{aligned} 
  \end{equation}
  where $C>0$ and $\varepsilon  >0$ are positive constants.  
\end{theorem}


\emph{Proof:}
Since the diffusion matrix is  $H = I + \delta A$, 
\begin{equation}\label{examER.1}
\begin{aligned}
  \mathbb{E}{H^2} =  \mathbb{E}{\left( {I + \delta A} \right)^2} \hfill  
   =  \delta^2 \mathbb{E}{A^2} + 2\delta \mathbb{E}A + I \hfill . \\ 
\end{aligned} 
\end{equation}
The expectation of $A$ can be computed directly as follows
\begin{equation}\label{examER.2}
\begin{aligned}
  \mathbb{E}A =
    b {\bold{1}^*}\bold{1} - b I \hfill . \\ 
\end{aligned} 
\end{equation}
Note that  when  $i \ne j$,
\begin{equation}\label{EA2.1}
		{\left[ {\mathbb{E}{A^2}} \right]_{i,j}} = \mathbb{E}\sum\nolimits_{k = 1}^n {A_{ki}^{}A_{kj}^{}}  = (n-2){b^2} \nonumber
	\end{equation}
represents the  expected value of  common neighbors between vertices  $i$ and $j$, and when   $i = j$,
\begin{equation}\label{EA2.2}
{\left[ {\mathbb{E}{A^2}} \right]_{i,i}} = \mathbb{E}\sum\nolimits_{k = 1}^n {A_{ki}^2}  = (n-1)b \nonumber
\end{equation}
represents  the average degree of vertex $i$. 
 Therefore, 
\begin{equation}\label{examER.3}
\begin{aligned}
   \mathbb{E}{A^2} =    (n-2){b ^2}{\bold{1}^*}\bold{1} + \left( {(n-1)b  - (n-2){b ^2}} \right)I \hfill .
\end{aligned} 
\end{equation}
For the convenience of calculation, considering that n represents the number of vertices, when $n$ is large enough, we make an approximation to the \eqref{examER.3} such that
\begin{equation}\label{examER.3aa}
\begin{aligned}
   \mathbb{E}{\hat{A}^2} =    n{b ^2}{\bold{1}^*}\bold{1} + \left( {nb  - n{b ^2}} \right)I \hfill .
\end{aligned} 
\end{equation}
Substituting \eqref{examER.2} and \eqref{examER.3aa} into \eqref{examER.1}, we have 
\begin{equation}\label{examER.H2}
\begin{aligned}
 \mathbb{E}{H^2} = \left( {n{b^2}{\delta ^2} + 2b\delta } \right){{\mathbf{1}}^*}{\mathbf{1}} + \left( {nb{\delta ^2} - n{b^2}{\delta ^2} - 2b\delta  + 1} \right)I. 
\end{aligned} 
\end{equation}
\ycaaa{For a matrix $X = a{{\mathbf{1}}^*}{\mathbf{1}}+bI  \in {\mathbb{R}^{n \times n}} $ with $a>0, b>0$, the $k$-sparse condition number can be calculated as 
\begin{equation}\label{condall1}
	\begin{aligned}
		\text{cond}(k,X) = \sqrt {k - (k - 1)\frac{{{b^2}}}{{n{a^2} + {b^2} + 2ab}}}. \end{aligned} 
	\end{equation}
Combining \eqref{examER.H2} and \eqref{condall1},  the $k$-sparse condition number of  $\mathbb{E}{H^2}$ can be upper bounded as
\begin{equation}\label{condall.2}
	\begin{aligned}
 {\text{cond}}(k,\mathbb{E}{H^2} ) \le  \sqrt k. 
\end{aligned} 
\end{equation}
}The inverse of $\mathbb{E}{H^2}$ is 
\begin{equation}\label{examER.H2ni}
\begin{aligned}
   {\left[ {\mathbb{E}{H^2}} \right]^{ - 1}} \hfill  
   =  {f_1}  \left( {I - {f_2} {{\mathbf{1}}^*}{\mathbf{1}}} \right) \hfill ,  \nonumber
\end{aligned} 
\end{equation}
where \begin{equation}\label{examER.f1}
\begin{aligned}
{f_1} 
\triangleq \frac{1}{{nb{\delta ^2} - n{b^2}{\delta ^2} - 2b\delta  + 1}}, \nonumber
\end{aligned} 
\end{equation}
and 
\begin{equation}\label{examER.f2}
\begin{aligned}{f_2}  \triangleq \frac{{n{b^2}{\delta ^2} + 2b\delta }}{{{n^2}{b^2}{\delta ^2} - n{b^2}{\delta ^2} + nb{\delta ^2} + 2nb\delta  - 2b\delta  + 1}}. \nonumber
\end{aligned} 
\end{equation}  
With a proper connection probability $ b$, when the number of vertices  $n$ is large,  ${f_2}  \to 0$, so we approximate ${\left[ {\mathbb{E}{H^2}} \right]^{ - 1}}$ as $ {1 /{f_1} }I$. Therefore the $k$-sparse condition number of ${\left[ {\mathbb{E}{H^2}} \right]^{ - 1}}$ can be approximate as
\begin{equation}\label{condall.3}
	\begin{aligned}
	 {\text{cond}}\left( {k,{{\left[ {\mathbb{E}{H^2}} \right]}^{ - 1}}} \right) \approx 1  . 
	\end{aligned} 
\end{equation}
Since $H_\mathcal{M}$ denotes a random selection of $m$ rows from $H$, it follows that   $\Gamma  = m{\left[ {\mathbb{E}H_\mathcal{M}^*H_\mathcal{M}^{}} \right]^{ - 1}} = n\mathbb{E}{\left[ {{H^*H}} \right]^{ - 1}}$. Therefore, combining  \eqref{condall.2} and \eqref{condall.3}, the $k$-sparse condition number of $\Gamma$ can be upper bounded as ${\kappa (\Gamma)}  \le \sqrt k $, and the  incoherence parameter ${\mu}$ is
\begin{equation}\label{examER.6}
\begin{aligned}
\mathop {  \max }\limits_{1 \le  i,j \le  n} \left\{ {\left| {h_{i,j}^{}} \right|} \right\} &\le 1  ,\\
 \mathop {\max }\limits_{1 \le  i,j \le  n} \left\{ {\left| {\left[ {H\Gamma } \right]_{i,j}^{}} \right|} \right\}& \le   \frac{1}{{{\delta ^2}\left( {b - {b^2}} \right)}}  .  \nonumber
\end{aligned} 
\end{equation}
Therefore, applying Theorem \ref{mainResult}, we can obtain the result in Theorem \ref{Lemma.er}. $\hfill\blacksquare$ 

 As illustrated  by Theorem \ref{Lemma.er},  for ER random networks, only   $\sim \log n$ samples are sufficient to ensure sparse signal recovery. 
In addition, the  number of samples  and the parameter of $\delta$ have a trend of reverse change.    This phenomenon 
 can be explained by the fact that a small value of $\delta$ leads to a diagonally dominant matrix $H$,  resulting in each vertex fusing less information from its neighbors and requiring more samples.
Moreover, the right-hand side of \eqref{erlem.1ll}
 is proportional to ${{ - \log \left( {b - {b^2}} \right)} \mathord{\left/
 {\vphantom {{ - \log \left( {b - {b^2}} \right)} {\left( {b - {b^2}} \right)}}} \right.
 \kern-\nulldelimiterspace} {\left( {b - {b^2}} \right)}}$ and reaches the minimum when $b=0.5$.
If $b$ tends to $0$ or $1$, all vertices of  the graph must be sampled to ensure recovery, i.e., $m=n$. 

\subsection{Small-world Network}
In this subsection, we consider the binary graph diffusion models on the small-world networks. The  small-world network is constructed by the Watts-Strogatz networks\cite{watts1998collective}, where the  adjacency matrix $A$  can be generated as follows\cite{porter2012small,rudolph2014algebraic}:
\begin{itemize}
	\item \textbf{Initial regular graph}:
	Each vertex is connected to its nearest  neighbors, forming a $d$-regular graph.  
	
	\item \textbf{Random  rewiring  part}: Rewire each connection randomly with probability $b $.  
\end{itemize}
The rewiring process is comprised of two  steps. Initially, each currently connected edge is independently erased with probability $b$. Subsequently, each edge pair is reconnected with probability $\frac{bd}{n-1}$, allowing for multiplicity. According to the construction rules of Watts-Strogatz network, we can get \cite{cai2017detection}
\begin{equation} \label{exa.0}
	\begin{aligned}
    \mathbb{P}\left( {{A_{ij}} = 1} \right) = \left\{ {\begin{array}{*{20}{c}}
  {1 - b\left( {1 - b\frac{d}{{n - 1}}} \right),}&{{\text{if }}0 < \left| {i - j} \right| < \theta,} \\ 
  {b\frac{d}{{n - 1}},}&{{\text{otherwise}}{\text{,}}} 
\end{array}} \right.\end{aligned} 
\end{equation}
where  $\theta \buildrel \Delta \over =   \frac{d}{2}\,\bmod \,{\kern 1pt} n - 1 - \frac{d}{2} $.     
Therefore,  the   expectation of  ${A}$ can be expressed as 
\begin{equation} \label{exa}
	\begin{aligned}
		 \mathbb{E}A = (1 - b)\left( {1 - b\frac{d}{{n - 1}}} \right)\mathbb{E}{A_{{\text{reg}}}} + b\mathbb{E}{A_{{\text{rand}}}}, 
	\end{aligned} 
\end{equation}
where the $A_{\text{reg}}$ and  $A_{\text{rand}}$ denote the adjacency matrix of  $d$-regular graph and the adjacency matrix of a random graph connected with probability $d/(n-1)$, respectively.  

 The following theorem  demonstrates that for small-world networks,   $	{\kappa (\Gamma)} $  has an upper bound
that is associated with the $d$-regular graph and the rewiring probability $b$.
To state the theorem, we need some definitions.
The parameter $\Delta _\kappa$ is defined as
\begin{equation} \label{sw.ks}
		\begin{aligned}
			 & \Delta _\kappa \triangleq  {\text{ }}{\delta ^2}{\left( {1 - b} \right)^2}\left( {1 - b\frac{d}{{n - 1}}} \right)^2\frac{{\left\| {A_{{\text{reg}},k}^4} \right\|_{1,1}^{1/2}}}{{{{\left\| {A_{{\text{reg}}}^2} \right\|}_{1 \to 2}}}} \\
 &\quad+ 2\delta \left( {1 - b} \right)\left( {1 - b\frac{d}{{n - 1}}} \right)\frac{{\left\| {A_{{\text{reg}},k}^2} \right\|_{1,1}^{1/2}}}{{\sqrt d }} + \sqrt k ,
\end{aligned} 
	\end{equation}
    where $A_{\text{reg},k}^2$ and $A_{\text{reg},k}^4$ is  the leading $k \times k$ principal submatrix of $A^2$ and $A^4$ as $A_{\text{reg},k}^2$ and $A_{\text{reg},k}^4$, respectively. Note that 
 $\left\| X \right\|_{1 \to 2}$ denotes the largest $L_2$-norm among all columns of $X$, and 
$ {\left\| X \right\|_{1,1}} = \sum\limits_{i,j} {\left| {{x_{ij}}} \right|} $  denotes the sum of absolute values of matrix elements of $X$. 
 
\begin{theorem} \label{Lemma.sw}
	For a small-world random network with $n$ vertices, the degree of each vertex is $d$ and the probability of rewiring an edge is $b$. Let $\alpha$ be a $k$-sparse vector diffused in the network by the binary  diffusion model $H = I+\delta A$, where  $A$ is the  unweighted adjacency matrix and $ 0<\delta   \le  1 $. Let $ \mathcal{M}$ be the sampling set as in Theorem \ref{mainResult}.     
 Then  with probability $ 1 - {e^{ - \varepsilon  }} - {3 \mathord{\left/
 {\vphantom {3 { n}}} \right.
 \kern-\nulldelimiterspace} { n}} $, the problem ($P_1$) has a unique minimizer  provided that
  \begin{equation} \label{erlem.1}
\begin{aligned}
 m \ge  C\left( {1 + \varepsilon } \right)k{\mu ^2} \cdot \Delta _\kappa \cdot \left( {\log n + \log \mu } \right), 
 \end{aligned} 
  \end{equation}
  where $C>0$ and $\varepsilon  >0$ are positive constants.   
\end{theorem}


\emph{Proof:}  
Similar to \eqref{examER.2}, we can obtain
\begin{equation} \label{exa.1}
	\begin{aligned}
		\mathbb{E}A_{{\text{rand}}}^{} = \frac{d}{n-1}{\mathbf{1}^*}\mathbf{1} - \frac{d}{n-1}I. 
	\end{aligned} 
\end{equation}	
We denote  $i$-th column of matrix ${\mathbb{E}A}$  as $\left[ {\mathbb{E}A} \right]_{: ,i}^{}$, which can be decomposed as
\begin{equation} \label{exa.2}
	\begin{aligned}
		\left[ {\mathbb{E}A} \right]_{: ,i}^{} =  \left( { 1- b} \right) \left( {1 - b\frac{d}{{n - 1}}} \right) {\left[ {{A_{{\text{reg}}}}} \right]_{:  ,i}}+\frac{{bd}}{n-1}  \mathds{1}_i,
	\end{aligned} 
\end{equation}
where $\mathds{1}_i \in {\mathbb{R}^n} $ is a hollow vector, with 0 in the $i$-th element and 1 in the others. For the matrix ${\mathbb{E}A_{}^2}$, when $i   \ne  j$, 
\begin{equation} \label{exa.3}
	\begin{aligned}
&  {\left[ {\mathbb{E}A_{}^2} \right]_{i,j}} = \left[ {\mathbb{E}A} \right]_{:,i}^*\left[ {\mathbb{E}A} \right]_{:,j}^{} \hfill \\
&   = {\left( {1 - b} \right)^2}{\left( {1 - b\frac{d}{{n - 1}}} \right)^2}{\left[ {A_{{\text{reg}}}^2} \right]_{i,j}} + {\left( {\frac{{bd}}{{n - 1}}} \right)^2}\left( {n - 2} \right) \hfill \\
 &  \quad + \left( {1 - b} \right)\left( {1 - b\frac{d}{{n - 1}}} \right)\frac{{bd}}{{n - 1}}\left( {\left[ {{A_{{\text{reg}}}}} \right]_{:,i}^* \mathds{1}_j + {\mathds{1}_i^*}{{\left[ {{A_{{\text{reg}}}}} \right]}_{:,j}}{\kern 1pt} } \right) \hfill \\
 &  = {\left( {1 - b} \right)^2}{\left( {1 - b\frac{d}{{n - 1}}} \right)^2}{\left[ {A_{{\text{reg}}}^2} \right]_{i,j}}    + f_1,  \hfill \\ 
\end{aligned} 
\end{equation}	
where 
\begin{equation} {f_1} \triangleq \left\{ {\begin{array}{*{20}{c}}
  {\frac{{\left( {n - 2} \right){b^2}{d^2}}}{{{{\left( {n - 1} \right)}^2}}} + \frac{{2b\left( {1 - b} \right)\left( {n - 1 - bd} \right)d\left( {d - 1} \right)}}{{{{\left( {n - 1} \right)}^2}}}},&{{\text{if }}\left( {i,j} \right) \in \mathcal{E},} \\ 
  {\frac{{\left( {n - 2} \right){b^2}{d^2}}}{{{{\left( {n - 1} \right)}^2}}} + \frac{{2b\left( {1 - b} \right)\left( {n - 1 - bd} \right){d^2}}}{{{{\left( {n - 1} \right)}^2}}}},&{{\text{otherwise}}{\text{.}}} 
\end{array}} \right. \nonumber
\end{equation}	  The second equation of \eqref{exa.3} is derived from \eqref{exa.2}, and the third equation of \eqref{exa.3} comes from the fact $\mathds{1}_i^*\mathds{1}_j= n-2$ and ${\left[ {{A_{{\text{reg}}}}} \right]_{:  ,i}^*}\mathds{1}_j={\mathds{1}_i^*}{\left[ {{A_{{\text{reg}}}}} \right]_{: ,j}}=d-1$ when $\left( {i,j} \right) \in \mathcal{E}$, otherwise  ${\left[ {{A_{{\text{reg}}}}} \right]_{:  ,i}^*}\mathds{1}_j={\mathds{1}_i^*}{\left[ {{A_{{\text{reg}}}}} \right]_{: ,j}}=d$ . 
When $i=j$, the   expectation of  ${A^2}$   represents  the average degree of vertex $i$, i.e., ${\left[ {\mathbb{E}A_{}^2} \right]_{i,i}} = d $.  Therefore, the  expectation of  ${A^2}$   can be expressed as
\begin{equation} \label{exa.4}
	\begin{aligned}
 \mathbb{E}A_{}^2 = {\left( {1 - b} \right)^2}{\left( {1 - b\frac{d}{{n - 1}}} \right)^2}A_{{\text{reg}}}^2 + {f_1}{{\mathbf{1}}^*}{\mathbf{1}} + {f_2}I,
	\end{aligned} 
\end{equation}	
where  ${f_2} \triangleq \left( {d - {{\left( {1 - b} \right)}^2}{{\left( {1 - b\frac{d}{{n - 1}}} \right)}^2}d - {f_1}} \right)$. By utilizing \eqref{exa}-\eqref{exa.4}, the expectation of  ${H^2}$  is  
  \begin{equation} \label{examsw.3}
\begin{aligned}
 & \mathbb{E}{H^2} = \mathbb{E}{\left( {I + \delta A} \right)^2} \hfill \\
  & = I + {\delta ^2}\mathbb{E}{A^2} + 2\delta \mathbb{E}A \hfill \\
  & = {\delta ^2}{\left( {1 - b} \right)^2}{\left( {1 - b\frac{d}{{n - 1}}} \right)^2}A_{{\text{reg}}}^2 \hfill \\
 &\quad  + 2\delta \left( {1 - b} \right)\left( {1 - b\frac{d}{{n - 1}}} \right){A_{{\text{reg}}}} + \Theta , \hfill \\ 
\end{aligned} 
  \end{equation}
where  $\Theta  \triangleq {f_3}{{\mathbf{1}}^*}{\mathbf{1}} + {f_4}I$  with $f_4>f_3>0$. According to \eqref{condall.2}, we can
upper bound   the $k$-sparse condition number of $\Theta$  as 
 ${\text{cond}}\left( {\Theta ,k} \right) \le  \sqrt k $.
 According to the properties of the $d$-regular graph,  we can  use the leading $k \times k$  principal submatrix  of ${A_{{\text{reg}}}^2}$ to obtain the  ${\lambda _{\max }}\left( {k,A_{{\text{reg}}}} \right)$, i.e., 
 \begin{equation} \label{examsw2.1}
 	\begin{aligned}
 	 {\lambda _{\max }}\left( {k,A_{{\text{reg}}}^{}} \right) = \frac{{\left\| {A_{{\text{reg}},k}^2} \right\|_{1,1}^{1/2}}}{{\sqrt k }}.
 \end{aligned} 
\end{equation}
The  ${\lambda _{\min }}\left( {k,A_{{\text{reg}}} } \right)$ can be calculated as
  \begin{equation} \label{examsw2.2}
 	\begin{aligned}
 {\lambda _{\min }}\left( {k,{A_{{\text{reg}}}}} \right) = \sqrt {\frac{d}{k}}  .
  \end{aligned} 
\end{equation}
 Combining \eqref{examsw2.1} and \eqref{examsw2.2}, the $k$-sparse condition number of the matrix ${A_{{\text{reg}}} }$ can be represented as
   \begin{equation} \label{examsw2.3}
 	\begin{aligned}
 {\text{cond}}\left( {k,A_{{\text{reg}}}^{}} \right) = \frac{{\left\| {A_{{\text{reg}},k}^2} \right\|_{1,1}^{1/2}}}{{\sqrt d }}.
   \end{aligned} 
\end{equation}
 Similarly, we can  use the leading $k \times k$  principal submatrix  of ${A_{{\text{reg}}}^4}$ to obtain the  ${\lambda _{\max }}\left( {k,A_{{\text{reg}}}^2} \right)$  
 \begin{equation} \label{examsw.4}
 	\begin{aligned}
{\lambda _{\max }}\left( {k,A_{{\text{reg}}}^2} \right) = \frac{{\left\| {A_{{\text{reg}},k}^4} \right\|_{1,1}^{1/2}}}{{\sqrt k }} .
\end{aligned} 
\end{equation}
 The  ${\lambda _{\min }}\left( {k,A_{{\text{reg}}}^2} \right)$ can be calculated as
\begin{equation} \label{examsw.5}
\begin{aligned}
 {\lambda _{\min }}\left( {k,A_{{\text{reg}}}^2} \right) = \frac{{{{\left\| {A_{{\text{reg}}}^2} \right\|}_{ 1 \to 2 }}}}{{\sqrt k }} .
\end{aligned} 
\end{equation}
Combining \eqref{examsw.4} and \eqref{examsw.5}, the $k$-sparse condition number of the matrix ${A_{{\text{reg}}}^2}$ can be represented as
\begin{equation} \label{examsw.6}
	\begin{aligned}
 {\text{cond}}\left( {k,A_{{\text{reg}}}^2} \right) = \frac{{\left\| {A_{{\text{reg}},k}^4} \right\|_{1,1}^{1/2}}}{{{{\left\| {A_{{\text{reg}}}^2} \right\|}_{1 \to 2}}}} .
\end{aligned} 
\end{equation}
Combining \eqref{examsw.3}, \eqref{examsw2.3} and \eqref{examsw.6}, we can obtain 
 \begin{equation} \label{examsw.7}
	\begin{aligned}
 &{\text{cond}}\left( {k,\mathbb{E}{H^2}} \right) \le  {\text{ }}{\delta ^2}{\left( {1 - b} \right)^2}\left( {1 - b\frac{d}{{n - 1}}} \right)^2\frac{{\left\| {A_{{\text{reg}},k}^4} \right\|_{1,1}^{1/2}}}{{{{\left\| {A_{{\text{reg}}}^2} \right\|}_{1 \to 2}}}} \\
 &+ 2\delta \left( {1 - b} \right)\left( {1 - b\frac{d}{{n - 1}}} \right)\frac{{\left\| {A_{{\text{reg}},k}^2} \right\|_{1,1}^{1/2}}}{{\sqrt d }} + \sqrt k  . 
\end{aligned} 
\end{equation}
For a non-negative vector $\alpha$ and a non-negative matrix $H$, according to Corollary \ref{lemm.sn},   
 ${\text{cond}}\left( {k,{{\left[ {\mathbb{E}{H^2}} \right]}^{ - 1}}} \right) \le  {\text{cond}}\left( {k,\mathbb{E}{H^2}} \right)$ holds.
Since $H_\mathcal{M}$ denotes a random selection of $m$ rows from $H$, it follows that   $\Gamma  = m{\left[ {\mathbb{E}H_\mathcal{M}^*H_\mathcal{M}^{}} \right]^{ - 1}} = n\mathbb{E}{\left[ {{H^*H}} \right]^{ - 1}}$. Therefore, we can draw the  conclusion in \eqref{sw.ks}. This completes the proof.
$\hfill\blacksquare$

 \ycaaa{ It can be demonstrated  that if the  rewiring  probability $b $ of the small-world network is equal to zero, the network is equivalent to a $d$-regular graph. If the  rewiring  probability $b $ is equal to one, the network is equivalent to an ER random graph. It can be seen from \eqref{erlem.1} that as the $b $   increases, the sparse condition number of $\Gamma$ will   decrease monotonically, and the incoherence  parameters ${\mu}^2$ also decrease  accordingly.  Consequently, the number of samples required for recovery will decrease as the  rewiring  probability $b$ of the small-world network increases. }

\section{Optimal sampling design with variable density}
In Section  III, we proved a sufficient condition for achieving unique recovery of the problem ($P_1$) \ycbbb{under uniform  random sampling strategy.}
The result  shows  that the sampling rate is related to ${\mu}^2$, which is the worst case of   $  {h_{i,j}^2} $ and  $  {\left[ { H   \Gamma } \right]_{i,j}^2} $. Currently, variable density sampling techniques are gaining popularity in compressed sensing as a means  of reducing sampling rates. Initially observed empirically in magnetic resonance imaging (MRI), variable density sampling aimed to speed up data acquisition by minimizing the amount of data collected \cite{mri}. Theoretical support for its effectiveness emerged in subsequent studies \cite{vary1,vary2,vary3} that  explored additional measurement constraints and structured sparsity patterns. This motivates us to  develop a sampling technique for diffused graph signals
on vertex  domain.
Specifically, we define two weights for each vertex $i$ as
\begin{equation}\label{op.1}
\begin{array}{l}
  { \phi  _i} \buildrel \Delta \over =   \mathop {\max }\limits_{j = 1,...,n} \left\{ {\left| {{h_{i,j}}} \right|} \right\}, \hfill \\
  {{\widetilde  \phi  }_i} \buildrel \Delta \over =   \mathop {\max }\limits_{j = 1,...,n} \left\{ {\left| {{{\left[ { H   \Gamma } \right]}_{i,j}}} \right|} \right\}  .\\ 
\end{array} 
\end{equation}
Then  we also define the sampling probability of each vertex $i$ as 
\begin{equation}\label{op.2}
    \begin{aligned}
 {p_i}{\rm{ }} \buildrel \Delta \over = \frac{{\sqrt {{\phi _i}{{\widetilde \phi }_i}} }}{{\sum\limits_{j = 1}^n {\sqrt {{\phi _j}{{\widetilde \phi }_j}} } }} .
    \end{aligned}
\end{equation}
It's obviously that   $\sum\nolimits_{i = 1}^n {{p_i}}  = 1$.
Now, we present a new sampling strategy and theoretical guarantee for unique recovery in following theorem.

\begin{theorem} \label{Result2}
		Consider the system \eqref{the3.1}, where $\alpha$ is the $k$-sparse input, $H$ the graph diffusion matrix. Let   $  \mathcal{M} = \{\omega_1, \ldots, \omega_m\} $ be a sampling set where $\omega_i$ is a independent 
	random variable with a distribution follows \eqref{op.2}.  With Assumption \ref{ass1} define the matrix $\Gamma$ as \eqref{Gamma}.   Then  with probability $ 1 - {e^{ - \varepsilon  }} - {3 \mathord{\left/
 {\vphantom {3 { n}}} \right.
 \kern-\nulldelimiterspace} { n}} $, the problem ($P_1$) has a unique minimizer  provided that
 \begin{equation} \label{res2.1}
\begin{aligned}
 m \ge  C\left( {1 + \varepsilon  } \right){\bar \phi}^2 k{\kappa (\Gamma)}\left( {\log n + \log \bar \phi } \right), 
 \end{aligned} 
  \end{equation}
  where $C>0$ and $\varepsilon  >0$ are positive constants, and the average weight $ \bar  \phi$  is
\begin{equation}\label{opp.2}
\begin{aligned}
 \bar  \phi   \buildrel \Delta \over = \frac{{\sum\limits_{j = 1}^n {  \sqrt{{ \phi  _i}{{\widetilde  \phi  }_i}} } }}{n}. 
\end{aligned}
\end{equation} 
\end{theorem}

The proof  of Theorem \ref{Result2} is given in Section VII.

It is noted that Theorem \ref{Result2} provides a performance guarantee at a lower sampling rate  compared to Theorem \ref{mainResult}. Theorem \ref{mainResult} relies on uniform   random sampling strategies where  the  worst-case  incoherence parameter  is replaced by the average incoherence parameter $  \bar { \phi  }$  in Theorem \ref{Result2}. As defined in \eqref{opp.2}, the $\bar { \phi}$ is always no greater than the   incoherence parameter $ {\mu} $.

\section{Proof of Theorem \ref{mainResult}}
In the proof of Theorem  \ref{mainResult}, we mainly adopt the technique  of \cite{unified}, which is mainly based on the golfing scheme. The golfing scheme has been widely used in matrix completion\cite{Gross}, structured matrix completion \cite{structured}, and compressed sensing\cite{ripless,kueng2014ripless}.
 To facilitate this proof, we denote the sparse support set of the sparse seed $\alpha$ as $ \mathcal{K} $, where  $\left| \mathcal{K} \right|  =  k$. The sparse support matrix ${D_\mathcal{K}} \in {\left\{ {0,1} \right\}^{n\times n}}$ is 
constructed as a diagonal matrix with diagonal elements defined as
\begin{equation} \label{CK}
\begin{aligned}
{\left[ {{D_\mathcal{K}}} \right]_{i,i}} \buildrel \Delta \over =  \left\{ {\begin{array}{*{20}{c}}
  1,&{{\text{if }}   i \in  \mathcal{K}  ,} \\ 
  0,&{{\text{otherwise}}{\text{.}}} 
\end{array}} \right.
\end{aligned} 
\end{equation}

Before starting the proof of Theorem \ref{mainResult}, we first introduce two Bernstein inequality lemmas related to random vectors and  matrices.

\begin{Lemma} \label{Lemma.vec}
(Vector Bernstein Inequality \cite{2012User}):
Let $v_i$  be a finite sequence of independent random vectors. 
Suppose that $ \mathbb{E} v_i = 0$ 
and  $\left\| {{{v}_i}} \right\| _2 \le  B $ almost surely for all $i$ and  $ \sum\nolimits_{i = 1}^n {\left\| {{v_j}} \right\|_2^2}  \le {\sigma ^2} $. 
Then for all  $0  \le  t  \le  {{{\sigma ^2}} \mathord{\left/
 {\vphantom {{{\sigma ^2}} B}} \right. \kern-\nulldelimiterspace} B} $, 
\begin{equation} \label{the5.1}
\begin{aligned}
 \mathbb{P}\left( {{{\left\| {\sum\limits_{i=1}^n {{v_i}} } \right\|}_2} \ge  t} \right) \le  \exp \left( { - \frac{{{t^2}}}{{8{\sigma ^2}}} + \frac{1}{4}} \right). 
\end{aligned} 
\end{equation}
\end{Lemma}
\begin{Lemma} \label{Lemma.matrix}
(Matrix Bernstein Inequality \cite{ripless}): Let  $\left\{ {X_i^{}} \right\} \in {\mathbb{C}^{d \times d}}$  be a finite sequence of independent random matrices. Suppose that  $\mathbb{E}X_i  = 0$  and  $\left\| {X_i^{}} \right\| _{2}   \le  B$  almost surely for all $i$ and
\begin{equation} \label{lema1.1}
\begin{aligned}
 \max \left\{ {{{\left\| {\sum\limits_{i = 1}^n {\mathbb{E}X_i^{}X_i^*} } \right\|}_2},{{\left\| {\sum\limits_{i = 1}^n {\mathbb{E}X_i^*X_i^{}} } \right\|}_2}} \right\} \le  {\sigma ^2}. 
\end{aligned}
\end{equation}
Then for all  $t  \ge  0$,  it holds that 
\begin{equation} \label{lema1.2}
\begin{aligned}
\mathbb{P}\left( {\left\| {\sum\limits_{i = 1}^n { X_i^{}} } \right\| _{2} \ge  t} \right)  \le  2d\exp \left( {\frac{{ - {t^2}/2}}{{{\sigma ^2} + Bt/3}}} \right).
\end{aligned}
\end{equation}
\end{Lemma}

\subsection{ Local Isometry }
\begin{Lemma}[Local Isometry] \label{lemm.li}  Let $\mathcal{K}$ be a fixed subset of $\mathcal{N}$ satisfying  $\left| {\cal K} \right| = k$. Let 
	$ \mathcal{M} = \{\omega_1, \ldots, \omega_m\}$ be a sampling set with $\omega_i$  being independent random variables with uniform distribution.  Then for each $t  >0 $,\\
\begin{equation} \label{lemLi.0}
\begin{aligned}
  &\mathbb{P}\left\{ {{{\left\| {D_\mathcal{K}^{}\left( {\frac{1}{m}\Gamma  {H_\mathcal{M}^*{H_\mathcal{M}}}   - I} \right){D_\mathcal{K}}} \right\|}_2} \ge  t} \right\} \hfill \\
  &  \le  2k\exp \left( { - \frac{{{{m{t^2}} \mathord{\left/
 {\vphantom {{m{t^2}} 2}} \right.
 \kern-\nulldelimiterspace} 2}}}{{{\mu ^2}k{\kappa (\Gamma)} + {{2{\mu ^2}kt} \mathord{\left/
 {\vphantom {{2{\mu ^2}kt} 3}} \right.
 \kern-\nulldelimiterspace} 3}}}} \right){\text{.}} \hfill \\ 
\end{aligned} 
\end{equation}
\end{Lemma}

\emph{Proof of Lemma \ref{lemm.li}: }
 The following
elementary bounds will be used repeatedly:
\begin{equation} \label{aa1}
\begin{aligned}
{\left\| {{D_\mathcal{K}} h_i^*} \right\|_2} \le  {\mu}  \sqrt { k} ,
\end{aligned}
\end{equation}
\begin{equation} \label{aa2}
\begin{aligned}
{\left\| {{D_\mathcal{K}} \Gamma  h_i^* } \right\|_2} \le {\mu}  \sqrt { k} .
\end{aligned}
\end{equation}

We begin with the proof for the case of sampling with replacement.  Define 
\begin{equation} \label{li.1}
\begin{aligned}
{M_i} \buildrel \Delta \over = {D_\mathcal{K}}\left(   {\Gamma  {h_{{  i}}^*}h_{{ i}}  - I} \right){D_\mathcal{K}}  ,{\text{  for all }}i = 1,...,m,
\end{aligned}
\end{equation}
where ${h_{i}} \in {\mathbb{R}^{  1 \times n  }}$  
represents the $i$-th row vector of matrix $H_{\mathcal{M}}$.
 Note that $\mathbb{E}   {{M_i}}  = 0$. The upper bound of $ \left\| {{M_i}} \right\|  _2  $ is 
\begin{equation} \label{li.2}
\begin{aligned}
  & {\left\| {{M_i}} \right\|_{2}} ={\left\| {{D_\mathcal{K}}  { \Gamma {h_{{  i}}^*} h_{{ i}}  } {D_\mathcal{K}}} \right\|_2} +1 \hfill \\
  & \le  {\left\| {{D_\mathcal{K}}\Gamma {h_{{  i}}^*}   } \right\|_2}\left\| { h_{{ i}}  
  {D_\mathcal{K}}} \right\|_2^{} + 1 \hfill  
   \le   2{{\mu}^2}  k \hfill, \nonumber
\end{aligned} 
\end{equation}
where the second inequality comes from \eqref{aa1} and \eqref{aa2}. For the upper bound of ${\left\| {\mathbb{E}M_i^*{M_i}} \right\|_2} $, we have
\begin{equation} \label{li.3}
\begin{aligned}
  & {\left\| {\mathbb{E}M_i^*{M_i}} \right\|_2} \hfill \\
   &= \left\| {\mathbb{E}{D_\mathcal{K}}\left( {h_i^*h{}_i\Gamma  - I} \right){D_\mathcal{K}}\left( {\Gamma h_i^*h{}_i - I} \right){D_\mathcal{K}}} \right\|_2^{} \hfill \\
  & = ||\mathbb{E}{D_\mathcal{K}}h_i^*h{}_i\Gamma {D_\mathcal{K}}\Gamma h_i^*h{}_i{D_\mathcal{K}} - \mathbb{E}{D_\mathcal{K}}\Gamma h_i^*h{}_i{D_\mathcal{K}} \\
   & \quad - \mathbb{E}{D_\mathcal{K}}h_i^*h{}_i\Gamma {D_\mathcal{K}} + {D_\mathcal{K}}||_2^{} \hfill \\
  & \le  {\left\| {\mathbb{E}{D_\mathcal{K}}\left( {h_i^*h{}_i\Gamma {D_\mathcal{K}}\Gamma h_i^*h{}_i - I} \right){D_\mathcal{K}}} \right\|_2} \hfill \\
   & \le  \max \left( {\left\| {{D_\mathcal{K}}\Gamma h_i^*} \right\|_2^2\left\| {\mathbb{E}{D_\mathcal{K}}h_i^*h{}_i{D_\mathcal{K}}} \right\|_2^{},1} \right) \hfill \\
  & \le  \max \left( {{\mu ^2}k\left\| {{D_\mathcal{K}}{\Gamma ^{ - 1}}{D_\mathcal{K}}} \right\|_2^{},1} \right) \hfill \\
  &  \le {\mu ^2}k{\kappa (\Gamma)}. \hfill  \nonumber
\end{aligned} 
\end{equation}
Similarly,
 $ {\left\| {\mathbb{E}{M_i}M_i^*} \right\|_2} 
   \le {\mu ^2}k{\kappa (\Gamma)}  $.
Since
\begin{equation} \label{li.5}
\begin{aligned}
\frac{1}{m}\sum\limits_{i = 1}^m {{M_i}}  = D_\mathcal{K}^{}\left( {\frac{1}{m}\Gamma {H_\mathcal{M}^*{H_\mathcal{M}}}- I} \right){D_\mathcal{K}}, \nonumber
\end{aligned} 
\end{equation}
applying the above results to Lemma \ref{Lemma.matrix}, we have 
\begin{equation} \label{li.6}
\begin{aligned}
 & \mathbb{P}\left\{ {{{\left\| {D_\mathcal{K}^{}\left( {\frac{1}{m}\Gamma 
 {H_\mathcal{M}^*{H_\mathcal{M}}}   - I} \right){D_\mathcal{K}}} \right\|}_2} \ge  t} \right\} \hfill \\
  & \le  2k\exp \left( { - \frac{{{{{m t^2}} \mathord{\left/
 {\vphantom {{{m t^2}} 2}} \right.
 \kern-\nulldelimiterspace} 2}}}{{{\mu}^2 k{\kappa (\Gamma)} + {{2{\mu}^2 kt} \mathord{\left/
 {\vphantom {{2{\mu}^2 kt} 3}} \right.
 \kern-\nulldelimiterspace} 3}}}} \right). \hfill  \nonumber   \\
\end{aligned} 
\end{equation}
$\hfill\blacksquare$

\subsection{Low Distortion}
\begin{Lemma} [Low Distortion] \label{lemm.ld}
	Let 
	$ \mathcal{M} = \{\omega_1, \ldots, \omega_m\}$ be a sampling set with $\omega_i$  being independent random variables with uniform distribution.  The $\alpha$ is a fixed vector and the $\mathcal{K}$ is a fixed subset of $\mathcal{N}$, where $\left| {\cal K} \right| = k$.  Then for each $0\le t   \le  {1 \mathord{\left/
 {\vphantom {1 2}} \right.
 \kern-\nulldelimiterspace} 2}$, the following inequality holds\\ 
\begin{equation} \label{the4.4}
\begin{aligned}
 & \mathbb{P}\left\{ {{{\left\| {D_\mathcal{K}^{}\left( {\frac{1}{m}\Gamma H_\mathcal{M}^* H_\mathcal{M} - I} \right){D_\mathcal{K}}\alpha } \right\|}_2} \ge  t{{\left\| {{D_\mathcal{K}}\alpha } \right\|}_2}} \right\} \\
 & \le \exp \left( { - \frac{{{mt^2}}}{{8{\mu}^2 k {\kappa (\Gamma)}}} + \frac{1}{4}} \right). 
\end{aligned} 
\end{equation}
\end{Lemma}

\emph{Proof of Lemma \ref{lemm.ld}: } Define 
\begin{equation} \label{the4a.1}
\begin{aligned}
{v_i} \buildrel \Delta \over = {D_\mathcal{K}}\left( { \Gamma  h_i^* h_i- I} \right){D_\mathcal{K}}\alpha ,{\text{  for all }}i = 1,...,m. 
\end{aligned}
\end{equation}
Then  ${v_i}$ satisfies $\mathbb{E} {v_i} = 0 $ and
\begin{flalign}\label{the4a.2}
  &\qquad {\left\| {{v_i}} \right\|_2} \hfill 
   =   {\left\| {{D_\mathcal{K}}\left( {\Gamma h_i^*{h_i} - I} \right){D_\mathcal{K}}\alpha } \right\|_2} \hfill \nonumber & \\
   &\qquad \le  {\left\| {{D_\mathcal{K}}\Gamma h_i^*} \right\|_2}\left\| {{h_i}{D_\mathcal{K}}} \right\|_2^{}\left\| {{D_\mathcal{K}}\alpha } \right\|_2^{} + {\left\| {{D_\mathcal{K}}\alpha } \right\|_2} \hfill  \nonumber &\\
    &\qquad \le  2{\mu ^2}k{\left\| {{D_\mathcal{K}}\alpha } \right\|_2}. \hfill  &   \nonumber
 \end{flalign}
\ycccc{For the upper bound of $\left\| {\mathbb{E}{v_i}} \right\|_2^2 $, we have}
\begin{equation} \label{the4a.3}
\begin{aligned}
  &\left\| {\mathbb{E}{v_i}} \right\|_2^2 \hfill  
   =  \mathbb{E}{\alpha ^*}{D_\mathcal{K}}\left( {h_i^*{h_i}H\Gamma  - I} \right){D_\mathcal{K}}\left( {\Gamma h_i^*{h_i} - I} \right){D_\mathcal{K}}\alpha  \hfill \\
   &\le  \mathbb{E}{\alpha ^*}{D_\mathcal{K}}h_i^*{h_i}\Gamma {D_\mathcal{K}}\Gamma h_i^*{h_i}{D_\mathcal{K}}\alpha  - {\alpha ^*}{D_\mathcal{K}}\alpha  \hfill \\
  & \le   {\mu ^2}k{\kappa (\Gamma)}\left\| {{D_\mathcal{K}}\alpha } \right\|_2^2. \hfill \nonumber
\end{aligned}
\end{equation}
Since
\begin{flalign}\label{the4a.5}
& \qquad \frac{1}{m}\sum\limits_{i = 1}^m {{v_i}}  = D_\mathcal{K}^{}\left( {\frac{1}{m} \Gamma  H_\mathcal{M}^*  H_\mathcal{M}  - I} \right){D_\mathcal{K}} \alpha, \nonumber &
\end{flalign}
applying the above results to Lemma \ref{Lemma.vec}, we have 
\begin{equation} \label{the4a.6}
\begin{aligned}
& \mathbb{P}\left\{ {{{\left\| {D_\mathcal{K}^{}\left( {\frac{1}{m} \Gamma  H_\mathcal{M}^*  H_\mathcal{M}  - I} \right){D_\mathcal{K}}} \alpha \right\|}_2} \ge  t  {\left\| {{D_\mathcal{K}}\alpha } \right\|_2}  } \right\} \\
 &\le   \exp \left( { - \frac{{m{t^2}}}{{8{\mu}^2 k{\kappa (\Gamma)}}} + \frac{1}{4}} \right).  \nonumber
\end{aligned} 
\end{equation}
$\hfill\blacksquare$ 

\subsection{Off-Support Incoherence}
\begin{Lemma}[Off-Support Incoherence]\label{lemm.osi}
	Let 
	$ \mathcal{M} = \{\omega_1, \ldots, \omega_m\}$ be a sampling set with $\omega_i$  being independent random variables with uniform distribution.  The $\alpha$ is a fixed vector and the $\mathcal{K}$ is a fixed subset of $\mathcal{N}$, where $\left| {\cal K} \right| = k$. Then for each $t > 0$, the following inequality  
\begin{equation} \label{l6.1}
\begin{aligned}
  \frac{1}{m}  {\left\| {D_{{\mathcal{K}^c}}^{} H _\mathcal{M}^*  H_\mathcal{M}   \Gamma {D_\mathcal{K}}\alpha } \right\|_ {\infty} }   <  t{\left\| {{D_\mathcal{K}}\alpha } \right\|_2} 
    \end{aligned} 
\end{equation}
holds except with probability
\begin{equation} \label{l6.2}
\begin{aligned}
 2n\exp \left( { - \frac{{m{t^2}/2}}{{{\mu}^2 {\kappa (\Gamma)} + {{{\mu}^2   \sqrt{k} t} \mathord{\left/
 {\vphantom {{{\mu}^2   \sqrt{k} t} 3}} \right.
 \kern-\nulldelimiterspace} 3}}}} \right), 
 \end{aligned} 
\end{equation}
where ${\mathcal{K}^c}$ represents the complement of ${\mathcal{K}}$.
\end{Lemma}

\emph{Proof of Lemma \ref{lemm.osi}: } Let $j \in \mathcal{K}^c$  be arbitrarily fixed. Define
\begin{flalign} \label{the6a.1}
 & \qquad  {r_i}\buildrel  \Delta \over =  \left\langle {{e_j}, h_i^* h_i   \Gamma {D_\mathcal{K}}  \alpha } \right\rangle ,{\text{  for all }}i = 1,...,m.& 
\end{flalign} 
Then  ${r_i}$ satisfies $\mathbb{E} {r_i} = 0 $ and
\begin{flalign*}  
  & \qquad \left| {{r_i}} \right| \le    \left| {e_j^* h_i^* h_i    \Gamma {D_\mathcal{K}}\alpha } \right| \hfill & \\
   & \qquad \le   \left| {e_j^* h_i^*    } \right|{\left\| {   h_i \Gamma {D_\mathcal{K}} } \right\|_2}{\left\| {{D_\mathcal{K}}\alpha } \right\|_2} \hfill & \\
   &\qquad \le   {\mu}^2  \sqrt{k}  {\left\| {{D_\mathcal{K}}\alpha } \right\|_2} \hfill.&  
\end{flalign*} 
\ycccc{For the upper bound of $ {\left\| {\mathbb{E}r_i^*{r_i}} \right\|_2} $, we have}
\begin{flalign*}  
&  \qquad {\left\| {\mathbb{E}r_i^*{r_i}} \right\|_2} \hfill = {\left\| {\mathbb{E}{\alpha ^*}{D_\mathcal{K}}\Gamma h_i^*{h_i}{e_j}e_j^*h_i^*{h_i}\Gamma {D_\mathcal{K}}\alpha } \right\|_2} \hfill & \\
&   \qquad  \le  {\left| {e_j^*h_i^*} \right|^2}{\left\| {\mathbb{E}{\alpha ^*}{D_\mathcal{K}}\Gamma h_i^*{h_i}\Gamma {D_\mathcal{K}}\alpha } \right\|_2} \hfill  & \\
  & \qquad \le  {\mu ^2}{\left\| {{D_\mathcal{K}}\Gamma {D_\mathcal{K}}} \right\|_2}\left\| {{D_\mathcal{K}}\alpha } \right\|_2^2 \hfill & \\
  &   \qquad \le   {\mu ^2}{\kappa (\Gamma)}\left\| {{D_\mathcal{K}}\alpha } \right\|_2^2. \hfill & 
\end{flalign*} 
Since 
\begin{flalign} 
&   \qquad \frac{1}{m}\sum\limits_{i = 1}^m {{r_i}}  = \frac{1}{m}\left\langle {{e_j},  H_\mathcal{M}^* H_\mathcal{M}   \Gamma {D_\mathcal{K}}\alpha } \right\rangle ,  &
\end{flalign}
applying the Lemma \ref{Lemma.matrix}  for
$d = 1$ and the union bound over all  $j \in   \mathcal{K}^c$ yields the claim.
 $\hfill\blacksquare$ 

\subsection{Dual Certification}
\begin{Lemma}  [Uniqueness by An Inexact Dual Certificate]  \label{Lemma.Uniq} Suppose that
    \begin{equation}\label{con.0}
   \left\| {D_\mathcal{K}^{}\left( {\frac{1}{m} \Gamma H_\mathcal{M}^* H_\mathcal{M}  - I} \right)D_\mathcal{K}^{}} \right\|_2 \le  \frac{1}{2}. 
    \end{equation}
    If there exists a vector $ \gamma $ in the row space of $H_\mathcal{M}  $ satisfying
 \begin{equation}\label{con.2}
 \begin{aligned}
{\left\| {D_\mathcal{K}^{}\left( {\gamma  - {\text{sgn}}\left( \alpha  \right)} \right)} \right\|_2} \le  \frac{1}{{7 {{\left\| { \Gamma  H } \right\|}_ {1 \to 2} {\left\| H \right\|}_ {1 \to 2} }}} 
\end{aligned}
\end{equation}
     and
    \begin{equation}\label{con.3}\begin{aligned}
     {\left\| { D_{{\mathcal{K}^c}}^{} \gamma } \right\|_\infty }  \le  \frac{1}{2} , \quad \quad  
     \end{aligned}\end{equation}
     then $\alpha$ is the unique minimizer to the problem ($P_1$).
\end{Lemma}

\emph{Proof of Lemma \ref{Lemma.Uniq}: }
    Let $ \hat \alpha  = \alpha + \beta  $  be the minimizer to \eqref{the3.1} and note that $ { H}_\mathcal{M} \beta  = 0$  since both $\alpha$ and $ \hat \alpha $ are feasible. To prove the claim, it suffices to show that $\beta= 0$. 
    
    Case 1: We first consider the case when $\beta$ satisfies
\begin{equation}\label{lemm4.1}
\begin{aligned}
   \left\| {D_\mathcal{K}^{}\beta} \right\| _2& \le  3 {\left\| \Gamma H \right\|_ {1 \to 2}  }{\left\| {{H }} \right\|_{1 \to 2} }{\left\| {D_{{\mathcal{K}^c}}^{}\beta} \right\|_2} \hfill . \nonumber
\end{aligned}
 \end{equation}
    We begin by observing that
    \begin{equation}\label{uniq.1}
     \begin{aligned}
  & {\left\| { \hat \alpha } \right\|_1} ={\left\| {D_\mathcal{K}^{}\alpha + D_\mathcal{K}^{}\beta} \right\|_1} + {\left\| {D_{{\mathcal{K}^c}}^{}\beta} \right\|_1} \\
 & \ge  {\left\| {D_\mathcal{K}^{}\alpha }  \right\|_1} + \left\langle {\operatorname{sgn} (D_\mathcal{K}^{}\alpha),D_\mathcal{K}^{}\beta} \right\rangle  + {\left\| {D_{{\mathcal{K}^c}}^{}\beta} \right\|_1}.
   \end{aligned}  
    \end{equation}
The second term on the right side of the inequality in \eqref{uniq.1} can be expressed as
\begin{equation}\label{uniq.2}\begin{aligned}
 & \left\langle {\operatorname{sgn} (D_\mathcal{K}^{} \alpha),D_\mathcal{K}^{}\beta} \right\rangle  \hfill \\
  & =  \left\langle {\operatorname{sgn} (D_\mathcal{K}^{}\alpha) - D_\mathcal{K}^{}\gamma ,D_\mathcal{K}^{}\beta} \right\rangle  + \left\langle {D_\mathcal{K}^{}\gamma ,D_\mathcal{K}^{}\beta} \right\rangle  \hfill \\
  & =  \left\langle {\operatorname{sgn} (D_\mathcal{K}^{}\alpha) - D_\mathcal{K}^{}\gamma ,D_\mathcal{K}^{}\beta} \right\rangle  - \left\langle {{D_{{\mathcal{K}^c}}}\gamma ,{D_{{\mathcal{K}^c}}}\beta} \right\rangle  \hfill , \nonumber  
\end{aligned} \end{equation}
where we used $\left\langle {D_\mathcal{K}^{}\gamma ,D_\mathcal{K}^{}\beta} \right\rangle  = \left\langle {\gamma ,\beta} \right\rangle  - \left\langle {{D_{{\mathcal{K}^c}}}\gamma ,{D_{{\mathcal{K}^c}}}\beta} \right\rangle  =  - \left\langle {{D_{{\mathcal{K}^c}}}\gamma ,{D_{{\mathcal{K}^c}}}\beta} \right\rangle$  and  $\left\langle {\gamma ,\beta} \right\rangle  =0$ since   $ \gamma$ is in the row space of $H$.
 Furthermore,
\begin{equation}\label{uniq.3}\begin{aligned}
  & \left| {\left\langle {D_\mathcal{K}^{}\gamma- \operatorname{sgn} (D_\mathcal{K}^{}\alpha),D_\mathcal{K}^{}\beta} \right\rangle +\left\langle {D_{{\mathcal{K}^c}}^{}\gamma ,D_{{\mathcal{K}^c}}^{}\beta} \right\rangle } \right| \hfill \\
   &  \le {\left\| {D_\mathcal{K}^{}\gamma  - \operatorname{sgn} (D_\mathcal{K}^{}\alpha)} \right\|_2}{\left\| {D_\mathcal{K}^{}\beta} \right\|_2} + {\left\| {D_{{\mathcal{K}^c}}^{}\gamma } \right\|_{ \infty  } }{\left\| {D_{{\mathcal{K}^c}}^{}\beta} \right\|_1} \hfill \\
    & \le     \frac{1}{{7n{{\left\| { \Gamma  H } \right\|}_{1 \to 2} }{{\left\| H \right\|}_{1 \to 2} }}} {\left\| {D_\mathcal{K}^{}\beta} \right\|_2} + \frac{1}{2}{\left\| {D_{{\mathcal{K}^c}}^{}\beta} \right\|_1} , \nonumber
\end{aligned}  \end{equation}
where the last inequality comes from \eqref{con.2} and \eqref{con.3}. Therefore,
\begin{equation}\label{uniq.4}\begin{aligned}
&  {\left\| { \hat \alpha  } \right\|_1} 
  \ge    {\left\| {{D_\mathcal{K}}\alpha } \right\|_1} - \frac{1}{{7 {{\left\| { \Gamma  H } \right\|}_{1 \to 2} }{{\left\| H \right\|}_{1 \to 2} }}}{\left\| {{D_\mathcal{K}}\beta} \right\|_2} + \frac{1}{2}{\left\| {{D_{{\mathcal{K}^c}}}\beta} \right\|_1} \hfill \\
    &\ge {\left\| {{D_\mathcal{K}}\alpha } \right\|_1} - \frac{3}{7}{\left\| {{D_{{\mathcal{K}^c}}}\beta} \right\|_2} + \frac{1}{2}{\left\| {{D_{{\mathcal{K}^c}}}\beta} \right\|_2} \hfill \\
   &\ge  {\left\| \alpha  \right\|_1} + \frac{1}{{14}}{\left\| {{D_{{\mathcal{K}^c}}}\beta} \right\|_2} \hfill, \nonumber
\end{aligned}  \end{equation}
 where the last inequality comes from  \eqref{lemm4.1} and ${D_{{\mathcal{K} }}}  \alpha = \alpha $. Then  $ {\left\| { \hat \alpha  } \right\|_1}  \le  {\left\| {  \alpha } \right\|_1}  \le  {\left\| { \hat \alpha  } \right\|_1} $. This implies ${{D_{{\mathcal{K}^c}}}\beta} =0$, which in turn implies  $\left\| {D_\mathcal{K}^{}\beta} \right\| _2=0$ by \eqref{lemm4.1}. Therefore, it follows that   $\beta= 0$.

Case 2: Next, we consider the complementary case when $\beta$ satisfies
 \begin{equation}\label{lemm4.8}
 \begin{aligned}
{\left\| {D_\mathcal{K}^{}\beta} \right\|_2}{\text{ }} > 3 {\left\| { \Gamma  H } \right\|_{1 \to 2} }{\left\| H \right\|_{1 \to 2} }{\left\| {D_{{\mathcal{K}^c}}^{}\beta} \right\|_2}{\text{ }}.
 \end{aligned}
     \end{equation}
     Since $\hat{\alpha}$  is feasible for \eqref{the3.3}. Thus  $H_\mathcal{M}\beta= H_\mathcal{M} \left( { \hat \alpha  - \alpha} \right) = 0$, which implies
\begin{equation}\label{lemm4.10}
\begin{aligned}
&  0 = \left\langle {{D_\mathcal{K}}\beta,\frac{1}{m} \Gamma  H _\mathcal{M}^*  H_\mathcal{M} \beta} \right\rangle  \hfill \\
   & = \left\langle {{D_\mathcal{K}}\beta,\frac{1}{m} \Gamma  H _\mathcal{M}^*  H_\mathcal{M} {D_\mathcal{K}}\beta} \right\rangle  + \left\langle {{D_\mathcal{K}}\beta,\frac{1}{m} \Gamma  H _\mathcal{M}^*  H_\mathcal{M}  {D_{{\mathcal{K}^c}}}\beta} \right\rangle . \hfill \\ 
\end{aligned} 
\end{equation}
The magnitude of the first term in the right-hand side of \eqref{lemm4.10} is lower-bounded by
\begin{equation}\label{lemm4.11}
\begin{aligned}
 & \left\langle {{D_\mathcal{K}}\beta,\left( { \frac{1}{m} \Gamma  H _\mathcal{M}^*  H_\mathcal{M} + I - I} \right){D_\mathcal{K}}\beta} \right\rangle  \hfill \\
  & = \left| {\left\langle {{D_\mathcal{K}}\beta,{D_\mathcal{K}}\beta} \right\rangle } \right| - \left| {\left\langle {{D_\mathcal{K}}\beta,\left( {I -  \frac{1}{m} \Gamma   H _\mathcal{M}^*  H_\mathcal{M}} \right){D_\mathcal{K}}\beta} \right\rangle } \right| \hfill \\
  & \ge  \left\| {{D_\mathcal{K}}\beta} \right\|_2^2 - {\left\| {I -  \frac{1}{m}{D_\mathcal{K}} \Gamma  H _\mathcal{M}^*  H_\mathcal{M} {D_\mathcal{K}}} \right\|_2}\left\| {{D_\mathcal{K}}\beta} \right\|_2^2 \hfill \\
  & \ge  \frac{1}{2}\left\| {{D_\mathcal{K}}\beta} \right\|_2^2, \hfill \\ 
\end{aligned} 
\end{equation}
where the last step follows from the assumption in \eqref{con.0}.
The second term in the right-hand side of \eqref{lemm4.10} is then upper bounded by
\begin{equation}\label{lemm4.15}
\begin{aligned}
  & \left\langle {{D_\mathcal{K}}\beta, \frac{1}{m} \Gamma  H _\mathcal{M}^*  H_\mathcal{M}{D_{{\mathcal{K}^c}}}\beta} \right\rangle  \hfill \\ 
  & \le   {\left\| { \frac{1}{m} \Gamma  H _\mathcal{M}^*  H_\mathcal{M} } \right\|_2}{\left\| {{D_\mathcal{K}}\beta} \right\|_2}{\left\| {{D_{{\mathcal{K}^c}}}\beta} \right\|_2} \hfill \\
   & \le    {\left\| { \Gamma  H } \right\|_{1 \to 2} }{\left\| H \right\|_{1 \to 2} }{\left\| {{D_\mathcal{K}}\beta} \right\|_2}{\left\| {{D_{{\mathcal{K}^c}}}\beta} \right\|_2} \hfill .\\ 
\end{aligned} 
\end{equation}
Applying \eqref{lemm4.11} and \eqref{lemm4.15} to \eqref{lemm4.10} provides
\begin{equation}\label{lemm4.16}
\begin{aligned}
 &  0\ge   \left| {\left\langle {{D_\mathcal{K}}\beta, \frac{1}{m} \Gamma  H _\mathcal{M}^*  H_\mathcal{M} {D_\mathcal{K}}\beta} \right\rangle } \right|
 - \left| {\left\langle {{D_\mathcal{K}}\beta, \frac{1}{m} \Gamma   H _\mathcal{M}^*  H_\mathcal{M}{D_{{\mathcal{K}^c}}}\beta} \right\rangle } \right| \hfill \\
   & \ge  \frac{1}{2}\left\| {{D_\mathcal{K}}\beta} \right\|_2^2 -  {\left\| { \Gamma  H } \right\|_{1 \to 2} }{\left\| H \right\|_{1 \to 2} }\left\| {{D_\mathcal{K}}\beta} \right\|_2 \left\| {{D_{{\mathcal{K}^c}}}\beta} \right\| _2 \hfill \\
  & \ge \frac{1}{2}\left\| {{D_\mathcal{K}}\beta} \right\|_2^2 - \frac{1}{3}\left\| {{D_\mathcal{K}}\beta} \right\|_2^2 \hfill  
   =   \frac{1}{6}\left\| {{D_\mathcal{K}}\beta} \right\|_2^2 \hfill \\
   &\ge  0,\hfill \nonumber
\end{aligned} 
\end{equation}
where the third inequality comes from \eqref{lemm4.8}. Then it is implied that ${{D_\mathcal{K}} \beta}=0$. By \eqref{lemm4.8}, we also have ${{D_{{\mathcal{K}^c}}} \beta}=0$. Therefore, ${ \beta}=0$, which completes the proof.
$\hfill\blacksquare$

\subsection{Existence of An Inexact Dual Certificate}
\begin{Lemma}[Existence of An Inexact Dual Certificate]\label{existence}
    With probability   $1 - {e^{ - \varepsilon  }} - {3 \mathord{\left/
 {\vphantom {3 n}} \right.
 \kern-\nulldelimiterspace} n}$, there exists a vector $ \gamma $ in the row space of $H_\mathcal{M}$ satisfying \eqref{con.2} to \eqref{con.3} when 
 \begin{equation} \label{lemm4.35}
\begin{aligned}
m \ge C \left( {1 +  \varepsilon  } \right){\mu}^2 k{\kappa (\Gamma)}\left( { \log n +  \log {\mu}} \right).
 \end{aligned} 
  \end{equation}
\end{Lemma}

\emph{Proof of Lemma \ref{existence}:}
Recall that the  $\mathcal{M} = \{\omega_1, \ldots, \omega_m\}$ is a sampling
set consists of i.i.d. random indices. We partition the multi-set $\mathcal{M}$ into $l$ multi-sets so that $\mathcal{M}_1$ consists of the first $m_1$ elements of $\mathcal{M}$, $\mathcal{M}_2$  consists of the next  $m_2$  elements of  $\mathcal{M}$, and so on, where  $\sum\nolimits_{i = 1}^l {{m_i} = m} $.

    The version of the golfing scheme in this paper generates a dual certificate $ \gamma   $ from intermediate vectors $b_i$  for $i = 0,...,l - 1$ by
\begin{equation}\label{lemm4.17}
\begin{aligned}
     \gamma  = \sum\limits_{i = 1}^l {\frac{1}{{{m_i}}} H _\mathcal{M}^* H_\mathcal{M}   \Gamma } {b_{i - 1}}, \nonumber
    \end{aligned} 
\end{equation}
where ${b_{i }}$ are generated as follows: first, initialize ${b_0} = \operatorname{sgn} \left( { \alpha } \right)$. Thus it follows that $\gamma$ lies in the row space of $H$ (since $H$ is symmetric). Next, generate $b_i$ recursively by
\begin{equation}\label{lemm4.18}
\begin{aligned}
{b_i} = {D_\mathcal{K}}\left( {I - \frac{1}{{{m _i}}}H _\mathcal{M}^* H_\mathcal{M}  \Gamma } \right){b_{i - 1}}. \nonumber
\end{aligned} 
\end{equation}
 The rest of the proof is devoted to show that $\gamma$ satisfies \eqref{con.2} and \eqref{con.3}. We show that $b_i$ satisfies the following two properties with high probability for each $i$. First,
\begin{equation}\label{lemm4.20}
\begin{aligned}
{\left\| {{b_i}} \right\|_2}  \le  {c_i}{\left\| {{b_{i - 1}}} \right\|_2}.  
\end{aligned} 
\end{equation} 
Second,
\begin{equation}\label{lemm4.21}
\begin{aligned}
 {\left\| {\frac{1}{{{m _i}}}{D_{{{\mathcal K}^c}}} 
  H _\mathcal{M}^* H_\mathcal{M}  \Gamma{b_{i - 1}}} \right\|_{  \infty } }   \le  {t_i}{\left\| {{b_{i - 1}}} \right\|_2} . 
\end{aligned} 
\end{equation} 
Let ${p_1}(i)$  (resp. ${p_2}(i)$) denote the probability that the inequality in \eqref{lemm4.20} (resp. \eqref{lemm4.21}) does not hold.
Since $b_{i-1}$ is independent of ${\mathcal{M}_i}$, by Lemma \ref{lemm.ld}, ${p_1}(i)$ is upper bounded by 
\begin{equation}\label{lemm4.22}
\begin{aligned}
 \exp \left( { - \frac{{{c_i^2}m_i}}{{8{{\mu}^2 } {k }  {\kappa (\Gamma)}  } }+ \frac{1}{4}}  \right). \nonumber
 \end{aligned} 
\end{equation}
Therefore,  $ {p_1}(i)  \le  \frac{1}{\tau  }{e^{ - \varepsilon  }} $ if 
\begin{equation}\label{lemm4.23}
\begin{aligned}
  {m _i}  \ge  \frac{{\left( {\varepsilon   + {1 \mathord{\left/
 {\vphantom {1 4}} \right.
 \kern-\nulldelimiterspace} 4} + \log \tau } \right)8{{\mu}^2 } {k }  {\kappa (\Gamma)}}}{{c_i^2}} . \nonumber
 \end{aligned} 
\end{equation}
On the other hand, since ${D_{\cal K}}{b_{i - 1}} = {b_{i - 1}}$  by Lemma \ref{lemm.osi}, ${p_2}(i)$ is upper bounded by
\begin{equation}\label{lemm4.25}
\begin{aligned}
2n\exp \left( { - \frac{{m_i t_i^2/2}}{{{\mu}^2 {\kappa (\Gamma)} + {\mu}^2 \sqrt k {t_i}/3}}} \right).
\end{aligned} 
\end{equation}
Therefore,  ${p_2}(i)  \le  \frac{1}{\tau  }{e^{ - \varepsilon  }}$ if
\begin{equation} \label{lemm4.26}
\begin{aligned}
  {m_i} \ge  2\left( {\log 2\tau   + \varepsilon   + \log n} \right){\mu}^2 {\kappa (\Gamma)}k\left( {\frac{1}{{kt_i^2}} + \frac{1}{{3{t_i}\sqrt k }}} \right) .
  \end{aligned} 
\end{equation}
We set the parameters similarly to the proof of \cite[Lemma 3.3] {ripless} as follows:
\begin{equation} \label{lemm4.26}
\begin{aligned}
 l &=  \left\lceil {{{\log }_2}\sqrt k   {{\left\| H \right\|}_{ 1 \to 2} }{{\left\| { \Gamma  H } \right\|}_{ 1 \to 2} }} \right\rceil  + 3,  \\
{c_i} &= \left\{ {\begin{array}{*{20}{l}}
  {1/\left\lceil {2\sqrt {\log n} } \right\rceil }, &\quad{\;\;\;{\kern 1pt} i = 1,2,3,} \\ 
  {1/2}, &\quad{\;\;\;{\kern 1pt} 3 < i  \le  l ,} 
\end{array}} \right. \hfill \\
 {t_i} &= \left\{ {\begin{array}{*{20}{l}}
  {1/\left\lceil {4\sqrt k } \right\rceil },&\quad\;{i = 1,2,3,} \\ 
  {{{\log n} \mathord{\left/
 {\vphantom {{\log n} {\left\lceil {4\sqrt k } \right\rceil }}} \right.
 \kern-\nulldelimiterspace} {\left\lceil {4\sqrt k } \right\rceil }}}, &\quad\;{3 <  i  \le  l ,} 
\end{array}} \right. \hfill \\
  {m _i} &=  \left\lceil {10(1 + \log 6 + \varepsilon  ){\mu}^2 {\kappa (\Gamma)}kc_i^{ - 2}} \right\rceil  . \\ 
\end{aligned} 
\end{equation}
By the construction of $ \gamma $, we have
\begin{equation} \label{lemm4.27}
\begin{aligned} 
 & {D_\mathcal{K}}\gamma   = \sum\limits_{i = 1}^l {\frac{n}{{{m _i}}}{D_\mathcal{K}} H _\mathcal{M}^* H_\mathcal{M}  \Gamma } {b_{i - 1}}{\text{ }} \hfill \\
  & = \sum\limits_{i = 1}^l {\left( {{D_\mathcal{K}}{b_{i - 1}} - {D_\mathcal{K}}\left( {I - \frac{1}{{{m _i}}} H _\mathcal{M}^* H_\mathcal{M}  \Gamma } \right){b_{i - 1}}} \right)} {\text{ }} \hfill \\
  & = \sum\limits_{i = 1}^l {\left( {{b_{i - 1}} - {b_i}} \right)}  
   = {b_0} - {b_l} \hfill  
    = {\text{sgn}}\left( \alpha  \right) - {b_l}{\text{ }} \hfill \\
  & = {D_\mathcal{K}}{\text{sgn}}\left( \alpha  \right) - {b_l}. \hfill    \nonumber
\end{aligned} 
\end{equation}
Therefore, \eqref{lemm4.20} implies
\begin{equation} \label{lemm4.28}
\begin{aligned}
    {\left\| {{D_\mathcal{K}}\left( {\gamma - \operatorname{sgn} \left( {\alpha} \right)} \right)} \right\|_2} \hfill 
  =   {\left\| {{b_l}} \right\|_2} \hfill  
    \le     \prod\limits_{i = 1}^l {{c_i}{{\left\| {\operatorname{sgn} \left( \alpha \right)} \right\|}_2}}  \hfill  
    \le    \frac{{\sqrt k }}{{{2^l}\log n}} \hfill . \nonumber
\end{aligned} 
\end{equation}
Next, by \eqref{lemm4.20} and \eqref{lemm4.21}, we have
\begin{equation} \label{lemm4.29}
\begin{aligned}
& {\left\| {{D_{{\mathcal{K}^c}}}\gamma } \right\|_\infty } \hfill 
   \le  \sum\limits_{i = 1}^l {{{\left\| {\frac{1}{{{m_i}}}{D_{{\mathcal{K}^c}}} H _\mathcal{M}^* H_\mathcal{M}  \Gamma {b_{i - 1}}} \right\|}_\infty}}  \hfill \\
   &  \le\sum\limits_{i = 1}^l {{t_i}{{\left\| {{b_{i - 1}}} \right\|}_2}}  \hfill  
   \le   \sqrt k \left( {{t_1} + \sum\limits_{i = 2}^l {{t_i}\prod\limits_{j = 1}^{i - 1} {{c_j}} } } \right). \hfill \\ 
\end{aligned} 
\end{equation}
By setting parameters as in \eqref{lemm4.26}, the right-hand side in \eqref{lemm4.29} is further upper bounded by
\begin{equation} \label{lemm4.30}
\begin{gathered}
\frac{1}{4}\left( {1 + \frac{1}{{2\sqrt {\log n} }} + \frac{{\log n}}{{4\log n}} +  \cdots } \right) < \frac{1}{2} . \nonumber
\end{gathered} \end{equation}
Then  we have shown that  $\gamma$  satisfies \eqref{con.3}. 

It remains to show that \eqref{lemm4.20} and \eqref{lemm4.21} hold with the desired probability. From \eqref{lemm4.23} and \eqref{lemm4.26}, it follows that
\begin{equation} \label{lemm4.31}
p_j(i)\leq\frac{1}{6}e^{-\varepsilon },\quad\forall i\in \{ 1,...,l\}  ,\forall j=1,2. \nonumber
 \end{equation}
In particular, we have 
\begin{equation} \label{lemm4.32}
\sum_{j=1}^2\sum_{i=1}^3p_j(i)\leq e^{-\varepsilon }. \nonumber
  \end{equation}
This implies that the first three  ${\mathcal{M}_i} $  satisfy  \eqref{lemm4.20} and \eqref{lemm4.21}  except with probability  ${e^{ - \varepsilon  }}$.
On the other hand, we also have
\begin{equation} \label{lemm4.33}
p_1(i)+p_2(i)<\frac13,\quad\forall i=4,\ldots,l.  \nonumber
  \end{equation}
In other words, the probability that ${\mathcal{M}_i}$ satisfies \eqref{lemm4.20} and \eqref{lemm4.21} is at least 2/3. The union bound doesn’t show that  ${\mathcal{M}_i}$ satisfies \eqref{lemm4.20} and \eqref{lemm4.21}  for all  $i  \ge  4$  with the desired probability.

As in the proof of \cite[Lemma3.3]{ripless }, we adopt the oversampling and refinement strategy by Gross \cite{Gross}. Recall that each random support set ${\mathcal{M}_i}$ consists of i.i.d. random indices. Thus ${\mathcal{M}_i} $ are mutually independent. In particular, we set ${\mathcal{M}_i} $  are of the same cardinality in \eqref{lemm4.26}. 
Therefore, ${\mathcal{M}_i} $ are i.i.d. random variables. We generate a few extra copies of ${\mathcal{M}_i} $ for  $i = l + 1,...,{l^{'}} + 3$  where ${l^{'}} = 3\left( {l - 3} \right)$ . 
Then  by Hoeffding’s inequality, there exists at least 
${{l^{'}} = 3\left( {l - 3} \right)}$ for 
$i  \ge  4 $ that satisfy \eqref{lemm4.20} and \eqref{lemm4.21} with probability  ${\text{1}} -  {3 \mathord{\left/ {\vphantom {1 n}} \right.
 \kern-\nulldelimiterspace} n} $.
(We refer to more technical details for this step to \cite[Sec. III.B]{ripless}. Therefore, there are $l$ good ${\mathcal{M}_i}s$ satisfying \eqref{lemm4.20} and \eqref{lemm4.21} with probability ${\text{1}} - {e^{ - \varepsilon  }} - 3/n$, and the dual certificate $v$ is constructed from these good ${\mathcal{M}_i} $. The total number of sampling for this construction requires
\begin{equation} \label{lemm4.34}
\begin{gathered}
 m \ge  40\left( {1 + \log 6 + \varepsilon  } \right){\mu}^2 k{\kappa (\Gamma)}\left( {3\log n + 3l} \right).  \nonumber
\end{gathered} 
  \end{equation}
It  can be simplified as
\begin{equation} \label{lemm4.35}
\begin{aligned}
 m \ge C\left( {1   + \varepsilon  } \right){\mu ^2}k{\kappa (\Gamma)}\left( {\log n + \log \mu } \right),  \nonumber
\end{aligned} 
\end{equation}
  where $C$ is a positive constant. $\hfill\blacksquare$

\section{Proof of Theorem \ref{Result2}}
  
Define 
\begin{equation}\label{opp.1}
\begin{aligned}
 \Phi   &\buildrel \Delta \over =  {\text{diag}}\left( { {{ \phi  _1}} ,..., {{ \phi  _n}} } \right) ,\\
  \widetilde{  \Phi }& \buildrel  \Delta \over = {\text{diag}}\left( { {{{\widetilde  \phi  }_1}} ,..., {{{\widetilde  \phi  }_n}} } \right) .
\end{aligned}
\end{equation}
Using the definition of $ \Phi$, $   \widetilde { \Phi } $ and $\bar { \phi  }$, we construct a pair of weighted transforms $\mathcal{H}_\mathcal{M}$ and $ \widetilde{\mathcal{H}}_\mathcal{M}$ as
\begin{equation}\label{opp.3}
\begin{aligned}
     \mathcal{H}_\mathcal{M} &\buildrel \Delta \over ={\bar{ \phi }} C_\mathcal{M}  H  \Phi ^{-1} ,
\\
 \widetilde {\cal H} _\mathcal{M} & \buildrel \Delta \over =  {\bar  \phi  } C_\mathcal{M} {H }\Gamma\widetilde{ \Phi }^{-1}  . 
\end{aligned}
\end{equation}
\ycccc{According the definition of $\mathcal{H}_\mathcal{M}$ and $\widetilde {\cal H} _\mathcal{M}$, we have}
\begin{equation}\label{opp.8}
 \begin{aligned}
 	{\left\| {{D_\mathcal{K}} \mathbf{h}_i^* } \right\|_2} \le &  \bar  \phi  \sqrt k ,\\
 {\left\| {{D_\mathcal{K}}  \widetilde{  \mathbf{h}}_i^*} \right\|_2}  \le& \bar  \phi  \sqrt k , 
 \end{aligned} \end{equation}
where  $\mathbf{h}_i$ and $\widetilde{  \mathbf{h}}_i$
represent  the $i$-th row vector of   $\mathcal{H}_\mathcal{M}$ and  $\widetilde {\cal H} _\mathcal{M}$, respectively.
Firstly, we focus on proving the local isometry property of variable density sampling. Define 
${M_i} \buildrel \Delta \over = {D_\mathcal{K}}\left(   {\widetilde { \mathbf{h}}_i^* \mathbf{h}_i  - I} \right){D_\mathcal{K}} $. The upper bound of  ${\left\| {{M_i}} \right\|_2} $ is 
\begin{equation}\label{oppa.1}
 \begin{aligned}
  & {\left\| {{M_i}} \right\|_2} ={\left\| {{D_\mathcal{K}}\left({  {  {\widetilde{\mathbf{h}}}_i^* } {\mathbf{h}_i}   - I} \right){D_\mathcal{K}}} \right\|_2}    \\
  & 
   \le   {\left\| {{D_\mathcal{K}} {  {\widetilde{\mathbf{h}}}_i^* }    } \right\|_2}\left\| { {  {\mathbf{h}}_i  } {D_\mathcal{K}}} \right\|_2^{} + 1 \hfill 
   \le  2{\bar \phi}^2 k \hfill. \nonumber
 \end{aligned} \end{equation}
Then referring to \eqref{li.3}, the upper bound of ${\left\| {\mathbb{E}M_i^*{M_i}} \right\|_2}$ can be replaced by
\begin{equation}\label{opp.9}
\begin{aligned}
 & {\text{ }}{\left\| {\mathbb{E}M_i^*{M_i}} \right\|_2} \hfill  
   =  \left\| {\mathbb{E}{D_\mathcal{K}}\left( {{\mathbf{h}}_i^*{\mathbf{\widetilde {h}}}_i^{} - I} \right){D_\mathcal{K}}\left( {{\mathbf{\widetilde {h}}}_i^*{\mathbf{h}}_i^{} - I} \right){D_\mathcal{K}}} \right\|_2^{} \hfill \\
  & = ||\mathbb{E}{D_\mathcal{K}}{\mathbf{h}}_i^{}{\mathbf{\widetilde {h}}}_i^*{D_\mathcal{K}}{\mathbf{\widetilde {h}}}_i^*{\mathbf{h}}_i^{}{D_\mathcal{K}}{\text{ }} - \mathbb{E}{D_\mathcal{K}}{\mathbf{\widetilde {h}}}_i^*{\mathbf{h}}_i^{}{D_\mathcal{K}} \\
   &\quad - \mathbb{E}{D_\mathcal{K}}{\mathbf{h}}_i^*{\mathbf{\widetilde {h}}}_i^{}{D_\mathcal{K}} + {D_\mathcal{K}}||_2^{} \hfill \\
   & \le \left\| {\mathbb{E}{D_\mathcal{K}}\left( {{\mathbf{h}}_i^{}{\mathbf{\widetilde {h}}}_i^*{D_\mathcal{K}}{\mathbf{\widetilde {h}}}_i^*{\mathbf{h}}_i^{} - I} \right){D_\mathcal{K}}} \right\|_2^{} \hfill \\
  & \le  \max \left( {\left\| {{D_\mathcal{K}}{\mathbf{\widetilde {h}}}_i^{}} \right\|_2^2\left\| {{D_\mathcal{K}}{\Gamma ^{ - 1}}{D_\mathcal{K}}} \right\|_2^{},1} \right) \hfill \\
 &  \le  {{\bar \phi }^2}k{\kappa (\Gamma)}. \hfill  \nonumber
\end{aligned}
\end{equation}
For the local isometry property, the following probability inequality holds
\begin{equation}\label{opp.10}
\begin{aligned}
 &\mathbb{P}\left\{ {{{\left\| {D_\mathcal{K}^{}\left( { \frac{1}{m}{\widetilde{ \mathcal{H}}_\mathcal{M}^*}\mathcal{H}_\mathcal{M} - I} \right){D_\mathcal{K}}} \right\|}_2} \ge  t} \right\}\\
 &  \le  {\text{ }}2k\exp \left( { - \frac{{{m t^2}/2}}{{{{\bar \phi }^2}k{\kappa (\Gamma)} + 2{{\bar \phi }^2}kt/3}}} \right). \nonumber
\end{aligned} 
\end{equation}
Next, we will low distortion property. Refer to \eqref{the4a.1} for the definition of ${v_i}$, define ${v_i} \buildrel \Delta \over =  {D_\mathcal{K}}\left( { \widetilde {\mathbf{h} }_i^* \mathbf{h} _i  - I} \right){D_\mathcal{K}}\alpha  $. The upper bound is 
\begin{equation}\label{oppa.1v}
\begin{aligned}
  &{\left\| {{v_i}} \right\|_2} \hfill  
   =  {\left\| {{D_\mathcal{K}}\left( { \widetilde {\mathbf{h} }_i^* \mathbf{h} _i - I} \right){D_\mathcal{K}}\alpha } \right\|_2} \hfill \\
  & \le   {\left\| {{D_\mathcal{K}} \widetilde {\mathbf{h} }_i^*   } \right\|_2}\left\| {\mathbf{h}_i{D_\mathcal{K}}} \right\|_2^{}\left\| {{D_\mathcal{K}}\alpha } \right\|_2^{} + {\left\| {{D_\mathcal{K}}\alpha } \right\|_2} \hfill \\
  &  \le 2{{\bar \phi }^2}k{\left\| {{D_\mathcal{K}}\alpha } \right\|_2} \hfill. \nonumber
\end{aligned} 
\end{equation}
Furthermore,
\begin{equation}\label{oppa.2}
\begin{aligned}
&  \left\| {\mathbb{E}{v_i}} \right\|_2^2 \hfill  
   =   \mathbb{E}{\alpha ^*}{D_\mathcal{K}}\left( {{\mathbf{h}}_i^*{\mathbf{\widetilde{h}  }}_i^{} - I} \right){D_\mathcal{K}}\left( {{\mathbf{\tilde h}}_i^*{\mathbf{h}}_i^{} - I} \right){D_\mathcal{K}}\alpha  \hfill  \\
  & \le   \mathbb{E}{\alpha ^*}{D_\mathcal{K}}{\mathbf{h}}_i^*{\mathbf{\widetilde{h}}}_i^{}{D_\mathcal{K}}{\mathbf{\widetilde{h}}}_i^*{\mathbf{h}}_i^{}{D_\mathcal{K}}\alpha  - {\alpha ^*}{D_\mathcal{K}}\alpha  \hfill \\
  &  \le {{\bar \phi }^2}k{\kappa (\Gamma)}\left\| {{D_\mathcal{K}}\alpha } \right\|_2^2. \hfill \nonumber
\end{aligned} 
\end{equation}
For the low distortion property, the following probability inequality holds
\begin{equation}\label{oppa.3}
\begin{aligned}
 &\mathbb{P}\left\{ {{{\left\| {D_\mathcal{K}^{}\left( { \frac{1}{m}  \widetilde {\mathcal{H}}_\mathcal{M}^* \mathcal{H}_\mathcal{M} - I} \right){D_\mathcal{K}}\alpha } \right\|}_2}
 \ge {{\left\| {{D_\mathcal{K}}\alpha } \right\|}_2}} \right\} 
 \\
  & \le  {\text{ }}\exp \left( { - \frac{{m{t^2}}}{{8{{\bar {\phi} }^2}k{\kappa (\Gamma)}}} + \frac{1}{4}} \right). \nonumber
\end{aligned} 
\end{equation}

Next, we will prove off-support incoherence. Refer to \eqref{the6a.1} for the definition of ${r_i}$, we have ${r_i}\buildrel  \Delta \over =   \left\langle {{e_j}, {\mathbf{h}}_i^*{\mathbf{\tilde h}}_i  {D_\mathcal{K}}\alpha } \right\rangle$.
Then  ${r_i}$ satisfies $\mathbb{E} {r_i} = 0 $ and
\begin{equation} \label{opp.13}
\begin{aligned}
  &\left| {{r_i}} \right| \hfill  
   \le   \left| {e_j^*  {\mathbf{h}}_i^*{\mathbf{\tilde h}}_i  {D_\mathcal{K}}\alpha } \right| \hfill \\
   &\le   {\left| {e_j^* {\mathbf{h}}_i^* } \right| }{\left\| {{D_\mathcal{K}}{\mathbf{\tilde h}}_i ^*  } \right\|_2}{\left\| {{D_\mathcal{K}}\alpha } \right\|_2} \hfill  \\
    & \le  {\bar{  \phi }}^2 \sqrt{k}{\left\| {{D_\mathcal{K}}\alpha } \right\|_2}. \hfill  \nonumber
\end{aligned} 
\end{equation}
Furthermore,
\begin{equation} \label{opp.14}
\begin{aligned} 
  & \mathbb{E}{r_i}r_i^* =   \mathbb{E}{\alpha ^*}{D_\mathcal{K}}{\mathbf{\widetilde{ h} }}_i^*{\mathbf{h}}_i^{}{e_j}e_j^*{\mathbf{h}}_i^*{\mathbf{\widetilde{ h}}}_i^{}{D_\mathcal{K}}\alpha  \hfill \\
 &  \le  {\left| {e_j^*{\mathbf{h}}_i^{}} \right|^2}\mathbb{E}{\alpha ^*}{D_\mathcal{K}}{\mathbf{\widetilde{ h}}}_i^*{\mathbf{\widetilde{ h}}}_i^{}{D_\mathcal{K}}\alpha {\text{ }} \hfill\\  
   & \le   {{\bar \phi }^2}{\kappa (\Gamma)}\left\| {{D_\mathcal{K}}\alpha } \right\|_2^2. \hfill \nonumber
\end{aligned} \end{equation}
Applying the above results to Lemma \ref{Lemma.matrix}  implies that
\begin{equation} \label{opp.15}
\begin{aligned}
 \left| {\left\langle {{e_j}, \mathcal{H}_\mathcal{M}^* \widetilde {\mathcal{H}}_\mathcal{M}{D_\mathcal{K}}\alpha } \right\rangle } \right| < t\left\| {D_\mathcal{K} \alpha } \right\|_2 , \nonumber
\end{aligned} 
\end{equation}
holds except with probability
\begin{equation}
 2n\exp \left( { - \frac{{m{t^2}/2}}{{{\bar  \phi }^2 {\kappa (\Gamma)} +{ \bar  \phi }^2 \sqrt k t/3}}} \right). \nonumber
\end{equation}
Referring to the subsequent proof of Theorem \ref{mainResult}, we can draw the conclusion of Theorem \ref{Result2}.

\section{Numerical Experiments}
\begin{figure*}[t] 
	\centering
	\subfloat[regular graph.]{
	\includegraphics[width=2.2in]{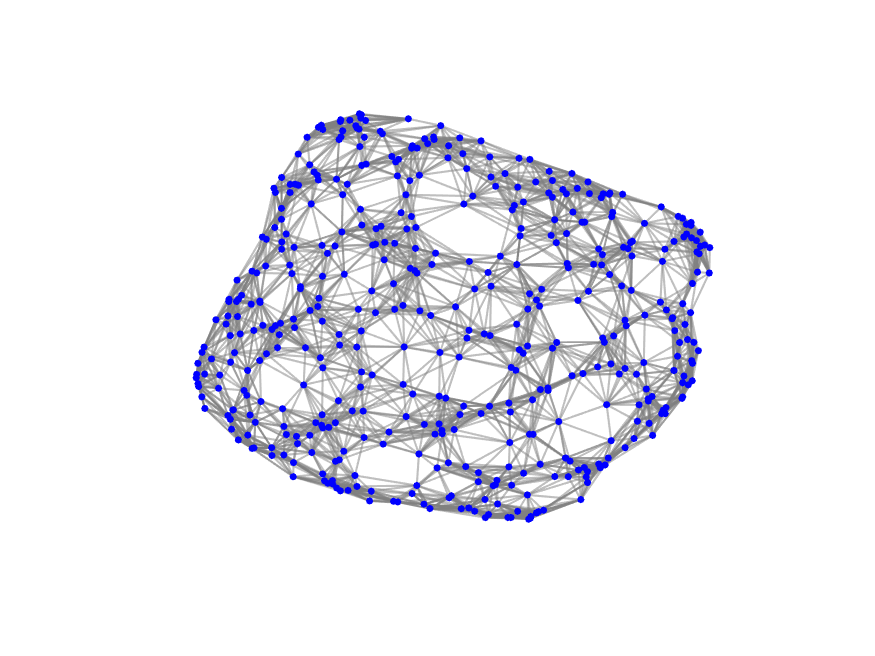}} 
   \quad  
    \subfloat[ER random graph.]{
	\includegraphics[width=2.2in]{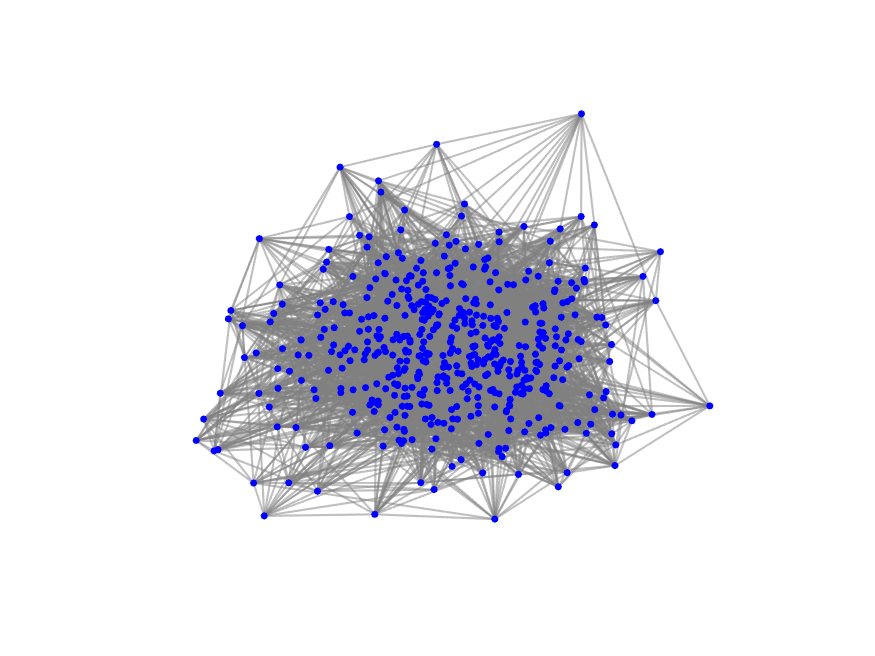}}
 \quad  
    \subfloat[star-like graph.]{
	\includegraphics[width=2.2in]{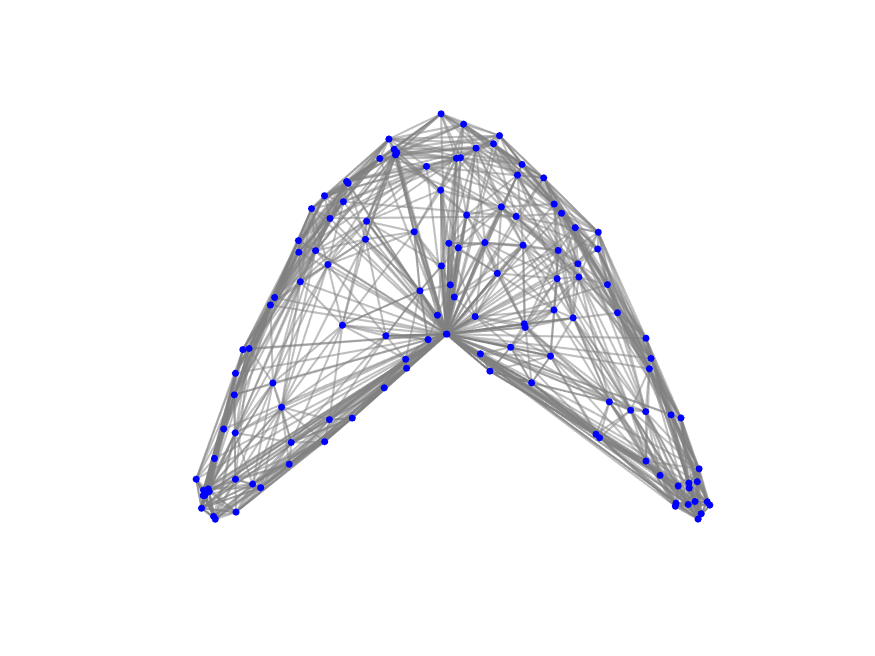}
         \label{starlike}}
	\caption{Three types of graphs.}
	\label{3graphSimu}
\end{figure*}

\begin{figure}[htbp]
\centering 
	\includegraphics[width=0.8\linewidth]{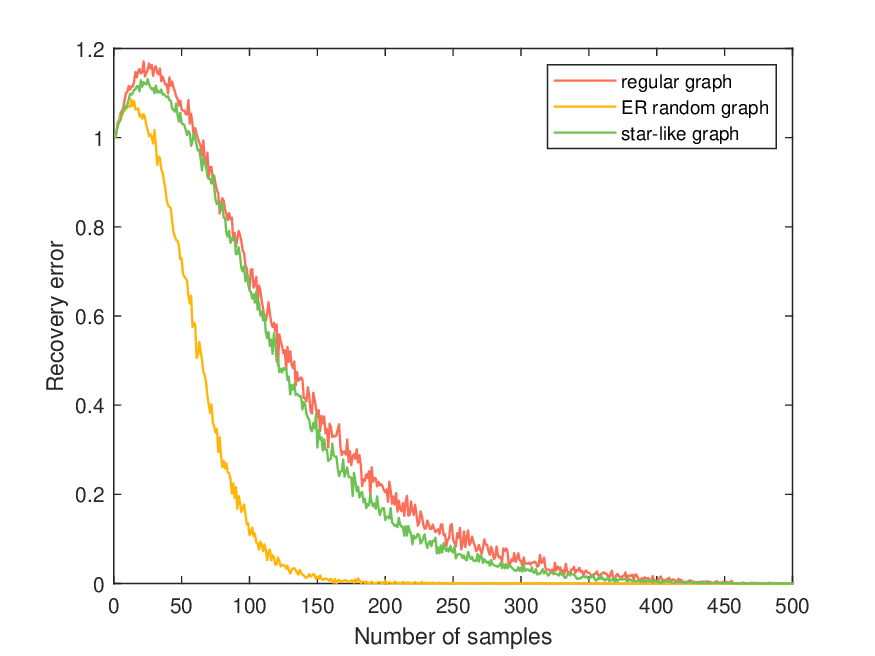}
	\caption{Comparison of sampling effects of several different types of graphs.  The horizontal axis represents the number of samples and the vertical axis indicates the normalized average recovery error computed by ${{{{\left\| {\alpha - {\hat \alpha }} \right\|}_2}} \mathord{\left/
 {\vphantom {{{{\left\| {\alpha  - {\hat \alpha }} \right\|}_2}} {{{\left\| {{\hat \alpha }} \right\|}_2}}}} \right.
 \kern-\nulldelimiterspace} {{{\left\| {{\hat \alpha }} \right\|}_2}}}$.}
 \label{erSenStar}
\end{figure}

  In this section, we present several simulation   experiments to show the effectiveness of the theoretical
results.   In Examples I-III, we demonstrate the effect of 
sampling performance under binary  diffusion models, which are described in Section IV with $\delta=1$. 
Specifically, different graphs with same number of vertices and same number of edges are tested in  Example I. The ER random networks with distinct  connection probabilities are tested in  Example II and small-world networks with distinct   rewiring  probabilities  are tested in  Example III.   We examine  the proposed variable
density sampling strategy in the case where the graph  diffusion models are generated by the doubly stochastic matrices in Example IV. 
All experiments are performed under different types of graphs available in the GSP toolbox\cite{2016GSPBOX}. 
We solve the  problem \eqref{the3.3} by performing the basis pursuit algorithms \cite{bp} to recover the sparse inputs for each experiment, and measure the accuracy of the recovery using the normalized average recovery error calculated  by ${{{{\left\| {\alpha - {\hat \alpha}} \right\|}_2}} \mathord{\left/
 {\vphantom {{{{\left\| {\alpha - {\hat \alpha}} \right\|}_2}} {{{\left\| {{\hat x}} \right\|}_2}}}} \right.
 \kern-\nulldelimiterspace} {{{\left\| {{\hat \alpha}} \right\|}_2}}}$.
 For each  number of samples, we do 500 iterations and average them.   
\yceee{\subsection*{\textbf{Example I:} Different Graphs with  Binary Diffusion Model}}
 This subsection presents   the sampling performance of  binary diffusion models for different graphs under random sampling strategies.  
  Three distinct  types of graphs were considered, each comprising the same number of vertices  and same number of edges:  the regular network, the ER random network, and the star-like network,  which are illustrated in Fig. \ref{3graphSimu}. 
  We refer to Fig. \ref{starlike} as a star-like graph where one vertex is connected to all other vertices.  
In this  experiment, the number of  vertices  is set as $n=501$, and the number of edges as $\left| \mathcal{E} \right| = 8417$  for all these three graphs. In addition, the sparse inputs  
$\alpha$ is randomly chosen  to be a vector with $k=4$ sparsity in each experiment. The recovery error  is shown in Fig. \ref{erSenStar}, and the relevant parameters of  ${\mu} $ and $ {\kappa (\Gamma)} $  involved in Theorem \ref{mainResult} are reported in Table \ref{tab1}.

It can be seen from Fig. \ref{erSenStar} that the ER random graph requires fewer samples to ensure accurate recovery compared to regular and star-like graphs, and the ${\mu} $ and ${\kappa_(\Gamma)} $ in the ER random graph are relatively small in Table \ref{tab1}. 
This fact  aligns with the theoretical results in Theorem \ref{mainResult}. Although the star-like graph displaying  larger incoherence parameters ${\mu}$ and sparse condition numbers $ {\kappa (\Gamma)} $ compared to the regular graph, they exhibit similar sampling performance.  This fact comes from  the star-like graph has only one fully connected vertex, and if this vertex does not belong to $\mathcal{K}$, the incoherence  parameter  ${\mu} $ and sparse condition number ${\kappa_(\Gamma)} $ are relatively small.

 
\begin{table}[htbp] 
\centering \caption{ \textbf{The relevant  parameters of three types of graphs.}}
\begin{tabular}{cccc}
\toprule[1.5pt]
type & regular graph  & ER random graph & star-like graph\\
\midrule[1pt]
 $ {\mu  } $     &  221.03 & 30.86 & 400.8\\
$ {\kappa (\Gamma)}$  &    8.58 & 2 & 271 \\
\bottomrule[1.5pt]
\label{tab1}
\end{tabular}
\end{table}
 
\subsection*{\textbf{Example II:} ER Random  Networks with  Binary Diffusion Model}

\begin{figure}[htbp]
\centering
	\includegraphics[width=0.8\linewidth]{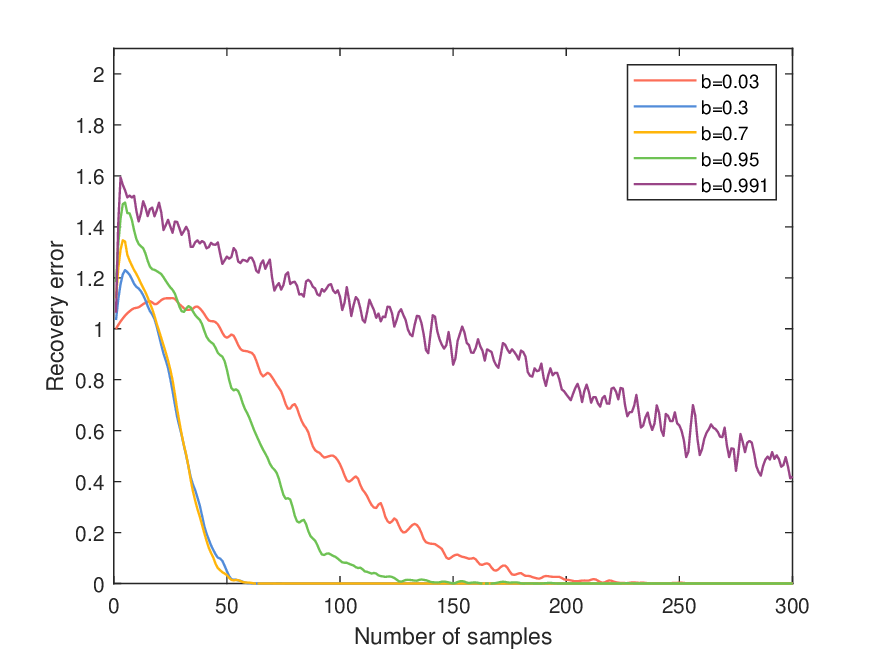}
	\caption{Comparison of sampling effects of ER random networks with different connection probabilities.}
 \label{ERP}
\end{figure}

In this experiment, the same binary diffusion model used in Example I was employed to evaluate the sampling and recovery performance of the ER random network under different connection probabilities.   We consider the large-scale  networks with a fixed number of vertices $n=10000$. The experiment results for the ER random networks with  connection probabilities $b =0.03,0.3, 0.7, 0.95, 0.991$ are presented in Fig. \ref{ERP}, and the  incoherence  parameters ${\mu}$ are given in Table \ref{tab2}. 
It can be seen from Table \ref{tab2} and Fig. \ref{ERP}  that  the number of samples  to ensure high probability recovery increases as  the incoherence  parameter  $\mu$ increases.  The result in \eqref{erlem.1} shows that the relationship between the number of samples  $m$   and the  connection probability $b $ is  $m \sim {{ - \log \left( {b - {b^2}} \right)} \mathord{\left/
 {\vphantom {{ - \log \left( {b - {b^2}} \right)} {\left( {b - {b^2}} \right)}}} \right.
 \kern-\nulldelimiterspace} {\left( {b - {b^2}} \right)}}$. Therefore, if $b $ tends to 0 or 1, the number of samples for recovery is relatively large. For an appropriate connection probability $b$, only samples proportional to the order of $\log n$ are needed to recover the signal with a high probability.
From the table \ref{tab2} and Fig. \ref{ERP}, it can be seen that when $b  = 0.3$ and $b  = 0.7$, the recovery performance  is relatively  close, which is consistent with our theoretical results ($m$ is symmetric around  $b=0.5$).  

\begin{table}[htbp] 
\centering \caption{\textbf{The relevant parameters of ER random networks with different connection probabilities.}}
\begin{tabular}{ccccccc}
\toprule[1.5pt]
$b$ & $ 0.03$ & $ 0.3$ & $ 0.7$ & $ 0.95$ & $ 0.991$ \\
\midrule[1pt]
 $ {{\mu} }$     & 34.36 & 4.76 & 4.76 & 21.05 & 112.12\\
\bottomrule[1.5pt]
\label{tab2}
\end{tabular}
\end{table}
\subsection*{\textbf{Example III:} Small-world Networks  with  Binary Diffusion Model}

\begin{figure}[htbp]
\centering
	\includegraphics[width=0.8\linewidth]{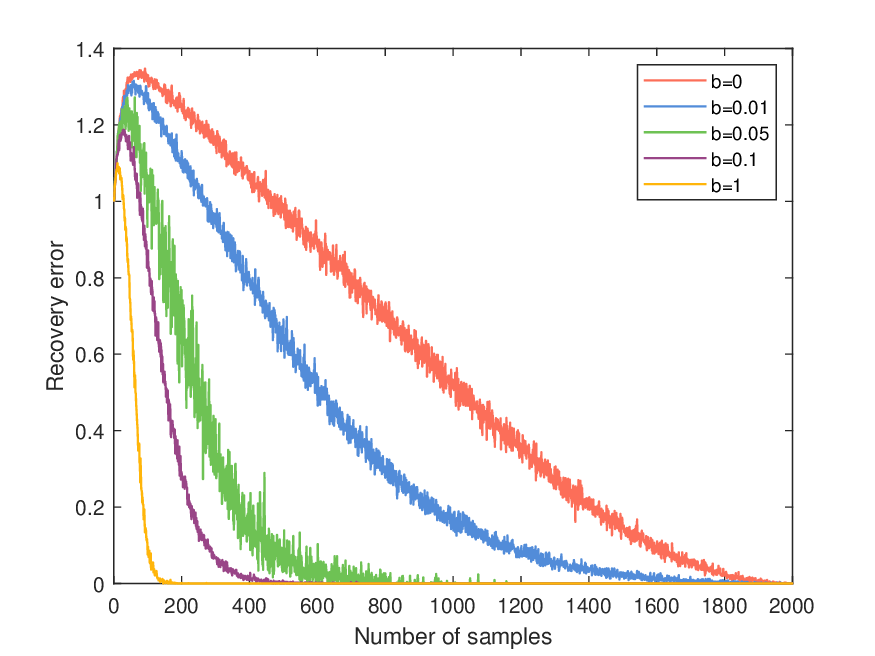}
	\caption{Comparison of sampling effects of small-world networks with different  rewiring  probabilities.}
 \label{WSFigure}
\end{figure}

Here, we test the sampling and recovery performance for the small-world networks with different  rewiring  probabilities $b $,
where the binary diffusion model is the same as in Example I. 
 In this experiment, we set the number of vertices  in  all involved networks to be $n=2001$ and the degree to be $d=41$. The experiment results for the small-world networks with $b =0, 0.01, 0.05, 0.1, 1$ are shown in Fig. \ref{WSFigure}. It is noted  from Section IV.B that due to the high degree of clustering of the regular network $A_{\text{reg}}$, the condition number $\kappa _s$ is relatively large.
 The Theorem \ref{Lemma.sw}  shows that  the condition number of $\mathbb{E}H_{\mathcal{M}}^*H_{\mathcal{M}}$ decreases as the  rewiring  probability $b $ increases. This result is supported by Fig. \ref{WSFigure}, which shows that  the number of samples  decreases as the  rewiring  probability increases.

\subsection*{\textbf{Example IV:} Diffusion Model with Doubly Stochastic Adjacency Matrices}  
\begin{figure}[htbp]
\centering
	\includegraphics[width=0.8\linewidth]{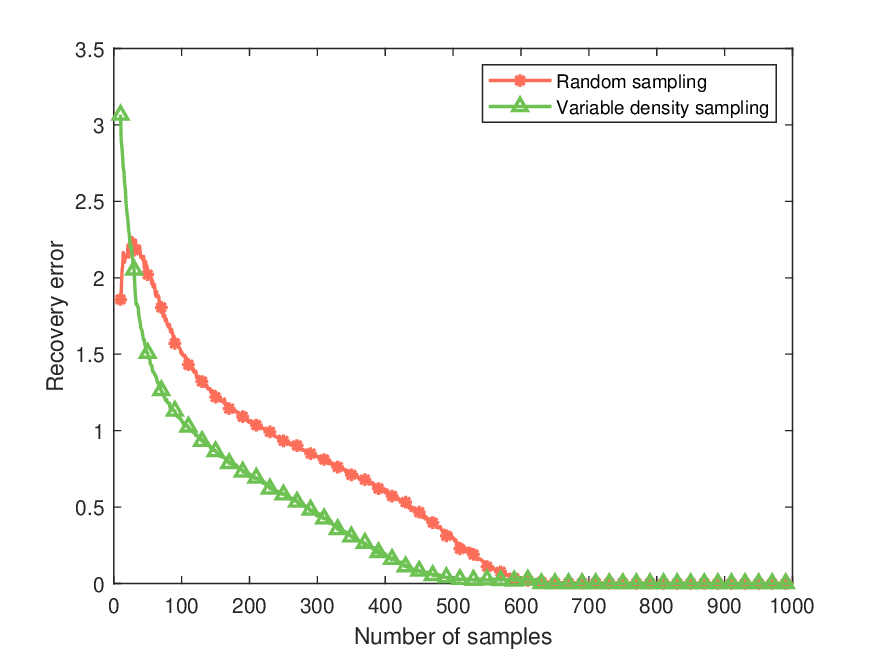}
	\caption{Comparison of sampling effects of the doubly stochastic matrix with different sampling probabilities.}
\label{communi}
\end{figure}

In this subsection, we evaluate the effectiveness of the variable density sampling proposed in Section V \ycrrr{when the diffusion matrix is a  doubly stochastic matrix, i.e., $H=A$,  $A\mathbf{1}= \mathbf{1} $ and $\mathbf{1}^* A= \mathbf{1} $.}  We conduct the experiments in a network with  $n=1000$ and we set the  graph  diffusion model as $H= A$, where $A$ is the doubly stochastic matrix, which is that are formed by the  Metropolis   rule \cite{metro}. The sparse inputs  
$\alpha$ for each experiment is randomly  set to be a  vector with sparsity $k=50$.
The result of the experiment is shown in Fig. \ref{communi}. It is clear  from Fig. \ref{communi}. that the proposed variable density sampling strategy can improve  the recovery performance to  some extent.

\section{CONCLUSION}

 This paper establishes sufficient conditions to guarantee recovery uniqueness  for uniform random sampling, providing  sampling guidelines  for a given diffusion model. 
 For specific binary diffusion models, a detailed analysis of the number of samples required for ER random networks and small-world networks has been conducted.
 In the case of ER random networks, it is demonstrated that a mere  $\sim \log n$ samples are sufficient to guarantee sparse signal recovery, and  how connection probability affects the number of samples required. For small-world networks, the number of samples required  and the  rewiring  probability tend to change inversely.
  Additionally, an adaptive sampling strategy is proposed to improve sampling performance by exploiting  variable density sampling technology from compressed sensing.   In future work,  we will   consider how to explore the network structure to design better recovery algorithms.

 

\bibliography{ref}
\bibliographystyle{ieeetr}

\newpage

 




\vfill

\end{document}